\documentclass{emulateapj}
\usepackage{apjfonts}
\usepackage{epsf}
\begin{document}
\slugcomment{}
\shortauthors{J. M. Miller et al.}
\shorttitle{Black Hole Disk Winds}

\title{Powerful, Rotating Disk Winds from Stellar-mass Black Holes}

\author{J.~M.~Miller\altaffilmark{1},
A. C. Fabian\altaffilmark{2}, 
J. Kaastra\altaffilmark{3,4}, 
T. Kallman\altaffilmark{5},
A. L. King\altaffilmark{6,7,8},
D. Proga\altaffilmark{9}, 
J. Raymond\altaffilmark{10}, 
C. S. Reynolds\altaffilmark{11}
}
 
\altaffiltext{1}{Department of Astronomy, University of Michigan, 500
Church Street, Ann Arbor, MI 48109-1042, USA, jonmm@umich.edu}

\altaffiltext{2}{Institute of Astronomy, University of Cambridge,
  Madingley Road, Cambridge CB3 OHA, UK}

\altaffiltext{3}{SRON Netherlands Institute for Space Research, Sorbonnelaan 2, 3584 CA Utrecht, NL}

\altaffiltext{4}{Department of Physics and Astronomy, Universiteit
  Utrecht, PO Box 80000, 3508 TA Utrecht, NL}

\altaffiltext{5}{NASA Goddard Space Flight Center, Code 662, Greedbelt, MD 20771, USA}

\altaffiltext{6}{Department of Physics, Stanford University, 382 Via
  Pueblo Mall, Stanford, CA, 94305}

\altaffiltext{7}{Einstein Fellow}

\altaffiltext{8}{Kavli Fellow}

\altaffiltext{9}{Department of Physics, University of Nevada, Las
  Vegas, Las Vegas, NV 89154, USA}

\altaffiltext{10}{Harvard-Smithsonian Center for Astrophysics, 60 Garden Street, Cambridge, MA 02138, USA}

\altaffiltext{11}{Department of Astronomy, University of Maryland, College Park, MD 20742-2421, USA}


\label{firstpage}

\begin{abstract}
We present an analysis of ionized X-ray disk winds observed in the Fe
K band of four stellar-mass black holes observed with {\it Chandra},
including 4U 1630$-$47, GRO J1655$-$40, H~1743$-$322, and GRS
1915$+$105.  High-resolution photoionization grids were generated in
order to model the data.  {\it Third-order} gratings spectra were used
to resolve complex absorption profiles into atomic effects and
multiple velocity components.  The Fe XXV line is found to be shaped
by contributions from the intercombination line (in absorption), and
the Fe XXVI line is detected as a spin-orbit doublet.  The data
require 2--3 absorption zones, depending on the source.  The fastest
components have velocities approaching or exceeding $0.01c$,
increasing mass outflow rates and wind kinetic power by orders of
magnitude over prior single-zone models.  The first-order spectra
require re-emission from the wind, broadened by a degree that is
loosely consistent with Keplerian orbital velocities at the
photoionization radius.  This suggests that disk winds are rotating
with the orbital velocity of the underlying disk, and provides a new
means of estimating launching radii -- crucial to understanding wind
driving mechanisms.  Some aspects of the wind velocities and radii
correspond well to the broad-line region (BLR) in active galactic
nuclei, suggesting a physical connection.  We discuss these results in
terms of prevalent models for disk wind production and disk accretion
itself, and implications for massive black holes in active galactic
nuclei.
\end{abstract}

\section{Introduction}
X-ray disk winds from low-mass X-ray binaries (LMXBs) are revealing
new facets of compact object accretion.  For instance, winds carry
away a significant fraction of the mass that is accreted onto the
compact object.  Estimates range from a few percent of the mass
accretion rate in the the inner disk, to several times the inflow rate
(see, e.g., King et al.\ 2013).  This means that mass transfer even in
LMXBs may be highly non-conservative in the very phase when the mass
transfer rate is expected to be highest.  This impacts binary
evolution models, and numerous specific predictions, including
e.g. the spin evolution of compact objects in LMXBs (for a review of
spins, see Miller \& Miller 2014; for a review of black hole X-ray
binaries, see Remillard \& McClintock 2006; also see Fragos \&
McClintock 2015).

Winds in stellar-mass black holes, in particular, may provide insights
into X-ray "warm absorbers" and faster outflows from Seyfert-1 active
galactic nuclei (AGN) and quasars.  Instrumental sensitivity curves
and intrinsic spectral shapes make the study of {\it low-ionization}
gas in warm absorbers relatively easy.  This readily-detected gas
could arise via irradiation of the "torus" (e.g. Kriss et al.\ 1996;
also see Lee et al.\ 2001); it may not directly probe the physics of
the inner accretion disk.  However, highly ionized components with
potentially different origins have emerged in deep exposures.  The
Chandra/HETG spectrum of MCG-6-30-15, for instance, contains strong Fe
XXV and Fe XXVI (He-like and H-like) absorption lines.  They are
blue-shifted by $v = 2000$~ km/s, require a column of $N_{\rm H}
\simeq 3\times 10^{23}~{\rm cm}^{2}$, and imply a kinetic power that
is about 10\% of the radiative luminosity (Young et al.\ 2005).  The
inferred launching radius is $10^{3-4}~ {\rm GM/c}^{2}$, putting this
component within or interior to the broad (emission) line region
(BLR).  These wind properties are remarkably similar to those measured
in, e.g., GRO J1655$-$40 (Miller et al.\ 2006, 2008; Kallman et
al.\ 2009; Neilsen \& Homan 2012).  Connections like this may underlie
emerging relationships between the kinetic power of winds and
accretion power, that span the black hole mass scale (King et
al.\ 2013).

Most importantly, perhaps, disk winds may reveal the fundamental
physics of disk accretion, making contact with simulations in a way
that continuum emission from the disk cannot.  Studies have shown that
disk winds and jets in X-ray binaries are anti-correlated (Miller et
al.\ 2006b, 2008; Neilsen \& Lee 2009; King et al.\ 2012, Ponti et
al.\ 2012).  There is actual evidence of absence: jets are truly
quenched in disk--dominated soft states (Russell et al.\ 2010), and
winds are not absent owing only to ionization effects (Miller et
al.\ 2012).  Viable explanations for this dichotomy include changes in
the inner accretion disk (perhaps related to the onset of an
advection-dominated accretion flow), and/or changes in dominant
magnetic field component above the disk (e.g. poloidal or toroidal).
More directly, some winds may be powered by magnetic forces, which
points to the underlying role of magnetic fields in mediating mass and
angular momentum transfer within the disk (Miller et al.\ 2006a,
2008).

Constraints on wind launching radii are central to understanding how
the winds are driven, and how much mass and power they carry.  This
requires photoionization modeling of the absorbing gas.  In most
cases, the gas density is not known, and radii must be deduced via $r
\leq L / N\xi$ (where $L$ is the ionizing luminosity, $r$ is the
launching radius, $N$ is the equivalent hydrogen column density, and
$\xi$ is the ionization parameter).  This assumes that $N = n r$
rather than $N = n \delta r$, and in this sense it is an upper limit.
In rare cases, the gas density can be measured directly, and wind
radii can be derived from $r = \sqrt(L/n\xi)$.  In the black hole GRO
J1655$-$40, the ratio of Fe XXII lines at 11.87~\AA~ and 11.92~\AA~
gives a density of $n \simeq 10^{14}~ {\rm cm}^{-3}$. (Miller et
al.\ 2006a, 2008).  In the black hole candidate MAXI J1305$-$704,
these line ratios imply a density of $n \simeq 10^{16-17}~ {\rm
  cm}^{-3}$ (though that wind may not escape the system; Miller et
al.\ 2014).  It is likely reasonable to assume that other winds with
numerous similarities to e.g. GRO J1655$-$40 may be equally dense, but
high line-of-sight column densities have prevented the detection of Fe
L-shell lines in prominent sources including 4U 1630$-$472,
H~1743$-$322, and GRS 1915$+$105.

Recent studies of a large population point toward an accretion disk
wind origin for the BLR in AGN (Tremaine et al.\ 2014).  Assuming that
Keplerian velocities dominate the width of these optical and UV lines,
radii of $10^{3-4}~{\rm GM/c}^{2}$ are obtained.  This range is commensurate
with wind radii inferred in stellar-mass black holes, again pointing
to a connection.  However, disk winds in black hole X-ray binaries
have not previously been detected in emission, preventing a similar
dynamical constraint on the launching radius.  This complicates most
attempts to understand the wind driving mechanism and the total mass
outflow rate, and hampers a clear association with stellar-mass disk
winds and the broad line region.

The best disk wind spectra have been modeled in an approximate
fashion.  Some analyses have attempted a line-by-line treatment with
Gaussian absorption functions (e.g. Miller et al.\ 2006b, Kubota et
al.\ 2007, Neilsen \& Lee 2009, Ueda et al.\ 2009).  Other efforts
have made use of single-zone absorption models generated with, e.g.,
XSTAR (e.g. Kallman \& Bautista 2001; Kallman et al.\ 2009) or Cloudy
(e.g. Ferland et al.\ 1998).  While this procedure is far more
physically self-consistent than line-by-line fitting, and while it has
likely captured the broad characteristics of disk winds correctly, it
has not yielded statistically acceptable fits.  It is likely that the
spectra contain additional information that has yet to be exploited.
Importantly, only one prior effort has included re-emission from the
wind (see Miller et al.\ 2014, concerning MAXI J1305$-$704), which
must occur whenever there is absorption.

In this work, we re-examine the best disk wind absorption spectra from
{\it Chandra}/HETG observations of stellar-mass black holes.  Our
analysis is restricted to the Fe K band, wherein the most highly
ionized gas -- likely originating closest to the black hole -- is
contained.  If stellar-mass black hole outflows are similar to Seyfert
warm absorbers, then this range also carries the bulk of the mass flux
(Crenshaw \& Kraemer 2012).  In order to best understand the atomic
and velocity structures present in the first-order HEG spectra, we
also consider the third-order spectra, which deliver three times
higher spectral resolution.  Indeed, the third-order spectra are only
a factor of a few lower in resolution than anticipated spectra from
{\it Astro-H} (Takahashi et al.\ 2010, 2014).  We employ a new version
of the XSTAR photoionization package to model both first- and
third-order data; its spectral resolution is 20 times higher than
prior versions, and it includes updated atomic data important to the
Fe K band.  Last, we allow for (re-)emission from the wind, with gas
parameters closely tied to corresponding absorption properties.

\section{Sources and Observations}
We require observations with extremely high sensitivity in the Fe K
band in first-order HETG spectra; this also ensures the best possible
sensitivity in third-order spectra.  However, for a combination of
astrophysical and pragmatic reasons, HETG observations of stellar-mass
black holes with strong signatures of disk winds are relatively rare.
Disk winds appear to be equatorial (Miller et al.\ 2006a,b; King et
al.\ 2012; also see Ponti et al.\ 2012), reducing the number of
sources in which these outflows can be observed.  Cygnus X-1 has been
observed extensively with the HETG, and it displays a complex
absorption spectrum (see, e.g., Schulz et al.\ 2002); however, this
source and other high-mass X-ray binaries are not considered as the
massive companion wind contaminates any signatures of a disk wind.
Cir X-1 was also omitted from consideration because its wind
features (Brandt \& Schulz 2000) are plausibly tied to its companion.

IGR J17091$-$3624 displayed a particularly fast and powerful disk wind
(e.g. King et al.\ 2012), but its low flux does not permit a deeper
analysis.  Modest sensitivity is also a general feature of neutron
star low-mass X-ray binaries; in the best cases, blue-shifted
absorption is sometimes detected (e.g. Ueda et al.\ 2004, Miller et
al.\ 2011), but blue-shifts are tentative in other sources and spectra
(e.g. 4U 1624$-$490; see Xiang et al.\ 2009).  The HETG spectra of MAXI
J1305$-$704 contain strong wind absorption features and even
re-emission at long wavelengths (Miller et al.\ 2014), but the
observation did not deliver reliable high-resolution spectra in the Fe
K region owing to a pointing offset.

The set of high-inclination black holes with very high flux levels and
multiple observations of disk winds is limited to four sources: 4U
1630$-$472, GRO J1655$-$40, H~1743$-$322, and GRS 1915$+$105.  Even
among these, not every spectrum has excellent sensitivity.  We
selected the observation of each source wherein the lines were best
defined (e.g. where the Fe XXV and Fe XXVI line fluxes divided by
their respective errors is maximal; this is distinct from the highest
column density, which would still give lines with large errors at low
flux or in a short expsoure).  For GRO J1655$-$40, H~1743$-$322, and
GRS 1915$+$105 this was possible using published measurements
(e.g. Miller et al.\ 2008; Miller et al.\ 2006b; Neilsen \& Lee 2009,
Ueda et al.\ 2009).  For 4U 1630$-$472, we made simple fits to
multiple spectra to determine the one with the highest sensitivity.
Table 1 lists the key properties of these observations, as well as an
observation of GX 339$-$4 that is examined later in this work.


Previous work on the selected spectrum of GRO J1655$-$40 already
points to a degree of complexity that is not captured by single-zone
photonionization models.  Miller et al.\ (2008) report that the best
single-zone model for the Fe K band in GRO J1655$-$40 -- model 1C in
Table 3 of that work -- only achieves a relatively poor fit:
$\chi^{2}/\nu = 2.88$.  Kallman et al.\ (2009) and Neilsen \& Homan
(2012) find evidence of multiple velocity components in the spectra of
GRO J1655$-$40.

The selected spectrum of H 1743$-$322 was not modeled with a
photoionzation code that has an XSPEC implementation, but Gaussian
models for the lines find inconsistent blue-shifts for the Fe XXV and
Fe XXVI lines ($v = 320\pm 160$ and $670\pm 170$~km/s, respectively;
Miller et al.\ 2006b).  This signals that a single absorbing zone is
not an adequate description of the spectrum.  Indeed, within Miller et
al.\ (2006b), it is noted that the observed Fe XXV and Fe XXVI lines
cannot arise in exactly the same gas.  A combination of zones with
different ionizations and velocities can certainly create this
disparity, however.

The shifts measured via single Gaussian models for the Fe XXV and Fe
XXVI lines in the selected observation of GRS 1915$+$105 are also
inconsistent (see Ueda et al.\ 2009).  The Fe XXV line shows a
red-shift, possibly indicating a contribution from the
intercombination line.  In contrast, the Fe XXVI line shows a
blue-shift.  This also signals a need for multiple velocity components.

At the time of writing, the selected spectrum of 4U~1630$-$472 has not
been described in detail using a photoionization code like XSTAR or
Cloudy.  However, it is included in a study that finds no evidence of
jet-based emission lines in this source (Neilsen et al.\ 2014; also
see Diaz-Trigo et al.\ 2013).

\section{Data Reduction}
All {\it Chandra}-specific data reduction and processing tasks were
performed using CIAO version 4.6, and corresponding calibration files.
Since the HEG has more collecting area than the MEG in the Fe K band,
we have only made use of HEG spectra in this work.  The {\it Chandra}
``tgcat'' facility (tgcat.mit.edu) provides calibrated first-order
spectra and associated response files for gratings observations.  We
therefore downloaded first-order HEG spectra and their corresponding
response files through ``tgcat''.

At the time of writing, complete higher-order data are not provided
through the ``tgcat'' facility, so full obserations were obtained from
the {\it Chandra} archive.  Calibrated spectral files are provided
through the archive, but response files must be generated by the user.
Response files for the third-order redistribution matrix files were
generated using the tool ``mkgrmf'', and these files were then used to
help construct ancillary response files using ``fullgarf''.

In order to achieve the best possible sensitivity, the CIAO tool
``add\_grating\_orders'' was used to combine opposing grating orders
into a single spectrum.  That is, the HEG$+$1 and HEG$-$1 spectra were
combined into a single first-order spectrum, and the HEG$+$3 and
HEG$-$3 spectra were combined into a single spectrum.  The
``add\_grating\_orders'' tool also adds the ancillary response files
to create a combined response files.  The gratings redistribution
matrix files are effectively the same for opposing orders.

We have relied on the HEG third-order data, as it provides the highest
possible spectral resolution.  However, it is worth noting that the
HEG second-order provides spectral resolution that is twice as high as
that available in first-order spectra.  Depending on the sensitivity
of the data, that resolution is only on the edge of being able to
separate the components of the Fe XXVI $Ly-a$ line, for instance.
Future work on these and other sources can make use of the HEG
second-order spectra.  In contrast, the MEG second-order spectra are
suppressed in the HETG, in order to facilitate order sorting.

All of the observations considered in this work were obtained by
running the ACIS-S array in ``continuous clocking'' mode, in order to
prevent photon pile-up.  In this mode, imaging information along the
narrow axis of the array is sacrificed in order to reduce the frame
time from 3.2~s to just 2.85~ms.  Since the ACIS chips are
medium-resolution spectrometers, accurate order sorting can still be
achieved despite the loss of imaging information.  The zeroth order
image lands on the S3 chip, and a sufficiently high incident flux
would cause frames from this chip to drop from the telemetry stream.
In each of the observations, then, a ``gray'' filter was applied
around the zeroth order position, so that only one in 10 or one in 20
incident zeroth order photons is recorded.  This enables tracking of
the aimpoint and the construction of the wavelength grid in the
dispersed spectra, while preventing frame loss.

This mode is capable of handling incredibly bright sources without
suffering photon pile-up.  If imaging information could be preserved,
source and background regions could be optimized to reduce background;
however, background is negligble for compared to the source flux in
these observations.  Aspects of this observational mode are also
described in detail in Miller et al.\ (2006b).

\section{Analysis and Results}

\subsection{Spectral Fitting Procedure}
All spectral fits were made using XSPEC version 12.8.1g (Arnaud et
al.\ 1996).  Fits to the {\it Chandra}/HETG data were made using the
Churazov weighting scheme (Churazov et al.\ 1996).  This procedure
weights channels by averaging the counts in surrounding channels.
Spectral regions with small errors are therefore weighted more
strongly than regions with large errors.  Minimization of the
$\chi^{2}$ fitting statistic then depends more strongly upon
well-defined lines than on relatively noisy parts of the spectrum.
The default weighting scheme was used in fits made to RXTE/PCA
spectra later in this work.

For dispersive spectrometers, it is natural to work in terms of
wavelength -- every bin has the same size, in units of wavelength.
However, the future of X-ray spectroscopic instrumentation,
particularly in the Fe K band, is the microcalorimeter.  The natural
unit for a calorimeter is energy, since every bin has the same size in
units of energy.  This analysis is intended to give the very best view
of accreting black holes as the field heads into the era of
calorimeters with {\it Astro-H} and {\it Athena}, so we have chosen to
present our analysis and results in energy units.  The data were
analyzed in their native bins and the results do not depend on whether
energy or wavelength units are adopted.

All fits to the first-order spectra of 4U 1630$-$472, GRO J16554$-$40,
H~1743$-$322, and GRS~1915$+$105 were made over the 5--10~keV range.
Fits to the third-order spectra of these black holes were made over
the 5--8~keV band, owing to a lack of signal and instrumental
residuals above 8~keV.  Where other sources and instrumental spectra
are considered, the fitting band is noted explicitly in the text.

\subsection{Initial Fits}

To give a view of potential structure and complexity in the Fe K band,
we fit each of the first-order spectra in the 5--10~keV band with a
phenomenological disk black body (``diskbb''; Mitsuda et al.\ 1984)
plus power-law model, modified by Galactic absorption fixed at
standard values.  This was accomplished using the ``tbabs'' model with
appropriate cross sections and abundances (Wilms, Allen, McCray 2000).
The power-law index was fixed to previously published values in each
case, where simultaneous RXTE observations had been made (GRO
J1655$-$40: $\Gamma = 3.5$, Miller et al.\ 2008; GRS 1915$+$105:
$\Gamma = 3.0$, Ueda et al.\ 2009; H~1743$-$322: $\Gamma = 2.4$,
Miller et al.\ 2006b).

Figure 1 shows the data/model ratios obtained when the {\it
  first-order} spectra are fit with these fiducial continua.
Blue-shifted Fe XXV and Fe XXVI absorption lines are the hallmarks of
accretion disk winds in stellar-mass black holes, and they dominate
the ratios.  However, on this relatively narrow wavelength range, at
least some of the lines are {\em not} simple.  The Fe XXV line in the
spectrum of H~1743$-$322 appears to be asymmetric, and the Fe XXVI
line may show an extended blue wing.  In 4U 1630$-$472, the Fe XXV
line shows some evidence of complexity, and the Fe XXVI line again
shows some possible structure.  The Fe XXV absorption in the spectral
ratio from GRS 1915$+$105 shows very clear structure, and the Fe XXVI
line again shows extension to high energy.  The ratio from GRO
J1655$-$40 may show the most absorption structure, with different
lines or velocity components potentially contributing to both Fe XXV
and Fe XXVI lines.

Beyond structure in the Fe XXV and Fe XXVI {\it absorption} lines,
this quick comparison of first-order spectra reveals the possibility
of {\it emission} lines.  It was previously noted that the absence of
strong, narrow emission lines indicates that the disk winds in these
high inclination sources must be equatorial: if the gas were
distributed in a more spherical manner, gas above the line of sight
would contribute {\it prominent} emission (Miller et al.\ 2006a,b).
However, for any plausible geometry, some re-emission from gas outside
of our line of sight should be detected.  Close scrutiny of this
narrow window around the Fe K region may indicate such emission, with
a structure that may be broadly consistent between the source spectra.
The putative emission is most prominent to the red of the strong
absorption lines; the ratios are flatter above 7~keV.  The structure
is at least qualitatively consistent, then, with the P-Cygni profiles
expected from an accretion disk wind (e.g. Dorodnitsyn \& Kallman
2009, Dorodnitsyn 2010, Puebla et al.\ 2011).

Figure 2 shows the data/model ratios obtained from the third-order
spectra of 4U 1630$-$472, H~1743$-$322, GRO J1655$-$40, and GRS
1915$+$105.  The spectra were fit with the same fiducial continuum
models used to characterize their corresponding first-order spectra.
Particularly in GRS 1915$+$105 and GRO J1655$-$40, it is apparant that
the Fe XXV line structure is indeed the result of multiple distinct
lines that are blurred together at first-order resolution.  This is
less distinct in 4U 1630$-$472, and unclear -- if present at all -- in
the third-order spectrum of H~1743$-$322.

Though often treated as a single line even at first-order HEG
resolution, the Fe XXVI line is actually a doublet owing to spin-orbit
coupling.  The expected separation of the two lines is about
$0.02~keV$ (6.952~keV and 6.973~keV, in a roughly 1:2 ratio of
oscillator strengths; see Verner, Verner, \& Ferland 1996).  This
splitting is revealed for the first time in GRO J1655$-$40.  Indeed,
one line pair is evident just below 7~keV, and a second just above
7.0~keV, signaling an absorption zone with an blue shift in excess of
0.01$c$.  The ratio of the line fluxes is inconsistent with 1:2; this
likely owes to saturation.

To quickly assess the significance of the lines in the putative
doublets in GRO J1655$-$40, we added Gaussian lines to the continuum
model.  A velocity width of 300 ${\rm km}~ {\rm s}^{-1}$ was assumed
as per Miller et al.\ (2008).  Dividing the Gaussian line flux by its
error gives one measure of line significance.  In the lower-velocity
pair, the lines are each significant at the $5\sigma$ level; in the
higher-velocity pair, the lines are each significant at the 3$\sigma$
level.  The spectra of 4U 1630$-$472 and H~1743$-$322 show weaker
evidence of having resolved the Fe XXVI doublet.  The splitting is not
evident in GRS 1915$+$105, but asymmetry is evident, again signaling
multiple outflow components at the highest possible resolution.

Figure 3 shows the first-order ratio spectra, plotted on top of the
higher-resolution third-order spectra.  The asymmetries and
complexities of the first-order line profiles generally correspond to
individual lines in the third-order spectra.  In short, the
third-order spectra contain a measure of precise information that can
be utilized to better understand the disk wind in these black holes.

In view of these results, we proceeded to: (1) model the third-order
spectra in detail to obtain additional constraints on gas properties
and velocities, and (2) combine third-order information and
corresponding emission components to model the sensitive first-order
spectra.  Both steps necessitated the development and application of
new photoionization models and fitting procedures, as described below.

\subsection{Photoionization grids and model implementation}
A sensible physical description of these spectra requires
photoionization modeling.  We generated large grids of synthetic
spectra using an update to a recent public version of the XSTAR
package (version 2.2.1bn19; e.g. Bautista \& Kallman 2001, Kallman et
al.\ 2009).  In order to take advantage of the third-order HEG
spectra, and in order to make optimal use of first-order spectra, two
specific improvements were made:

$\bullet$ Prior versions of XSTAR only allowed for 10,000 spectral
bins; however, this is insufficient for comparison to third-order HETG
spectra.  For instance, this resolution blends the Fe XXVI 
doublet into a single feature.   The synthetic spectra in our grids
were all generated using 200,000 spectral bins.  This reveals all of
the lines in full detail and is suited to the resolution of the
third-order spectra.

$\bullet$ Improved atomic data for He-like Fe XXV intercombination
lines was included.  Prior versions of XSTAR under-estimated the
strength of this feature, driving fits toward false velocity shifts
and ionization parameters in order to model the Fe XXV intercomination
line in terms of Fe XXIV absorption.  Fits with the updated version of
XSTAR used in this work correctly account for the intercombination
line in the Fe XXV complex, and deliver more accurate gas parameters.
Line wavelengths and oscillator strengths for the fine struture
components of the He-like and H-like ions in the updated version of
XSTAR were taken from the NIST database
(physics.nist.gov/PhysRefData/ASD/lines\_form.html).  Only the updates
to the intercombination line data are important for our analysis.

Table 2 lists the critical input parameters used to generate XSTAR
grids using the ``xstar2xspec'' functionality.  The XSPEC fitting
package recognizes the grids as models and can interpolate between
grid points to find the gas parameters that best correspond to the
observed spectra.  For each source, the grids spanned a range of $3
\leq {\rm log}(\xi) \leq 6$ and $21 \leq {\rm log(N)} \leq 23.8$.  All
grids were generated assuming a gas density of $n = 10^{14}~{\rm g}~
{\rm cm}^{-3}$, drawing upon the cases where the density has been
measured directly.  Whenever possible, parameter values -- especially
disk temperature and source luminosity -- were taken from values in
the literature.  Solar abundances were used for all sources apart from
GRS 1915$+$105, wherein an enhanced iron abundance has sometimes been
reported (e.g. Lee et al.\ 2002; XSTAR takes abundances from Grevesse,
Noels, \& Sauval 1996).  A covering fraction of $\Omega/4\pi = 0.2$
was used for GRO J1655$-$40 based on prior work (Miller et al.\ 2008);
a value of 0.5 was used for the other sources, consistent with King et
al.\ (2013).

Though the observed spectra are best characterized by a combination of
disk blackbody plus power-law models, the XSTAR grids were generated
using only a simple blackbody function with an equivalent temperature.
This is a simplification that introduces a degree of error and
inconsistency.  Model 1c in Table 3 of Miller et al.\ (2008) provides
the most useful point of comparison, for evaluating the degree to
which measurements are skewed by ignoring the power-law component.
The prior work uses a lower-resolution photoionization grid,
containing older atomic data.  However, comparing values of column
density, ionization, velocity, mass outflow rate, and kinetic power
derived Miller et al.\ (2008) to those found in the simplest fits
later in this work, the maximal systematic error induced by ignoring a
power-law component may be a factor of $\simeq$2 in column density and
ionization, but only 50\% in radius, mass outflow rate, and kinetic
power, since differing values partly compensate for each other.  The
true level of systematic error is likely lower, since numerous factors
changed between the 2008 analysis and this paper, and some of those
must also contribute to the differences.

The ``xstar2xspec'' script produces three distinct files for potential
inclusion in XSPEC modeling: a multiplicative table model
(``xout\_mtable.fits'') that imprints absorption onto a continuum, an
emission model that is the line spectrum emitted in all directions as
a result of the initial absorption (``xout\_ain.fits''), and an
emission component that represents the emission lines transmitted into
the pencil beam through which the absorption is seen
(``xout\_aout.fits'').  The last component is expected to be
negligible, and was not included in our models.

It may be possible to construct geometries and viewing angles that
would cause the observed re-emission from an absorbing wind to sample
a different set of gas properties than probed by the bulk of the
absorption.  However, assuming that the properties of the emitting gas
are the same as the absorbing gas permits a degree of simplicity.
Thus, in all spectral fits presented in this work, the equivalent
neutral hydrogen column density of the gas ($N_{\rm H}$) and the
ionization parameter ($\xi$) were linked, creating distinct zones with
characteristic gas properties for both the absorbing and emitting gas.

The emission component carries a flux normalization parameter: ${\rm
  K} = f {\rm L}_{tot,38}/d_{kpc}^{2}$, where $f = \Omega/4\pi$ is the
covering fraction, ${\rm L}_{tot,38}$ is the total luminosity of the
source in units of $10^{38}~ {\rm erg}~ {\rm s}^{-1}$, and $d_{kpc}$
is the distance in units of kpc.  A value of unity would indicate that
each parameter has been estimated with excellent accuracy, but modest
fractional errors in one or more parameters are likely, and can lead
to large deviations from unity.  In particular, distances within the
Milky Way are often particularly uncertain, and luminosity estimates
can vary depending on the spectral model assumed and the energy band
over which the flux is measured.  In our fits, then, the value of each
emission component was loosely bounded: $0.1 \leq {\rm K} \leq 10$.
This range acknowledges plausible uncertainties while demanding a
minimum contribution from the emission component.

Basic assumptions about the gas geometry further informed the manner
in which the spectral models were constructed.  The absorption lines
are clearly blue-shifted; else they could not be described as disk
winds.  The photoionized absorption components were therefore bounded
from below to have either zero velocity shift, or a blue shift.  The
photoionized emission lines are likely come from the full cylinder in
which the disk wind arises; in this case, velocities should cancel and
the emission should have zero net velocity shift (but a significant
velocity width).  If the far side of the cylinder (with respect to the
central engine) contributes preferentially for any reason, the
emission would have a net red-shift.  Figures 1 and 3 suggest that any
emission is strongest to the red of the (blue-shifted) absorption
components.  Based on these expectations, we bounded the emission
components to have a red-shift between zero and the absolute value of
the observed blue-shift in the same component (e.g. $v_{emis.} \leq
-1.0\times v_{abs.}$).  Note that when these conditions were relaxed,
emission components were not found to be blue-shifted, and absorption
components were not found to be red-shifted.  Rather, this step merely
permitted a degree of simplicity in constructing models.

A particular detail of the photoionization model implementation had to be
determined via numerous fitting experiments.  A priori, it is not
clear if the photoionized emission spectrum from points around the
wind cylinder might also be affected by local or ambient wind
absorption.  Numerous trials revealed that vastly improved fits are
obtained when the emission spectrum is not obscured.  

Last, we allowed for a narrow Gaussian emission line at 6.4~keV,
consistent with emission from neutral or nearly-neutral Fe, to
describe any remaining illumination of distant or ambient cold gas.
Reflection from the outer disk, for instance, might be one means by
which such a feature could be generated.  It is unlikely that any such
emission would be closely tied to the wind, and including this
Gaussian allows the XSTAR models to fit ionized wind features, as intended.

For a single zone, then, the full continuum plus photoionization
model was implemented into XSPEC in the following manner:\\

\noindent {\footnotesize $tbabs\times((abs\times (gauss+diskbb+powerlaw) + emis)$.}\\

Models for the spectra requiring 2--3 zones are implemented as follows:\\

\noindent {\footnotesize $tbabs\times((abs_{1}\times abs_{2}\times (gauss+diskbb+powerlaw) + emis_{1} + emis_{2})$.}\\

Further fits explored potential broadening of the photoionized emission
components, using smoothing functions to model velocity broadening of
the line spectrum.  Simple Gaussian broadening was explored using the
``gsmooth'' model, bounding the maximum smoothing to 0.2~keV (at
6~keV) with an index of $\alpha = 0$ (smoothing is not a function of energy):\\

\noindent {\footnotesize $tbabs\times((abs\times (gauss+diskbb+powerlaw) + gsmooth\times emis)$.}\\

Broadening with a line function that includes Doppler shifts and
illumination effects was implemented using the ``rdblur'' function,
which is a convolution model based on the ``diskline'' function
(Fabian et al.\ 1989).  This function is not technically appropriate
for spinning black holes.  However, Schwarzschild and Kerr metrics are
very similar far from the black hole, and this analytic blurring model
has the advantge of easily extending to large radii.  An outer radius
of $10^{5}~ r_{g}$ was fixed in all cases.  An emissivity index of $q
= -3$ was used ($J \propto r^{q}$ in ``rdblur''), appropriate for
large distances from a corona (see, e.g., Wilkins \& Fabian 2012).
The inclination parameter within ``rdblur'' was constrained based on
limits derived for the inner disk, in order to allow for a flared wind
geometry.  The final model was implemented exactly as per
``gsmooth'':\\

\noindent {\footnotesize $tbabs\times((abs\times (gauss+diskbb+powerlaw) + rdblur\times emis)$.}\\

\hspace{0.1in}

\subsection{Fits to the third-order spectra}

Figures 4--7 show fits to the third-order spectra of GRO J1655$-$40,
GRS 1915$+$105, 4U 1630$-$472, and H~1743$-$322, approximately ordered
from most complex to least complex.  Each spectrum is shown with the
best-fit model given in Tables 3--7 (models 1655-3a, 1915-3a, 1630-3a,
1743-3a).  The spectra were ``unfolded'' in a manner that removes the
instrumental effective area curve, without falsely imprinting the
model upon the data (e.g. the XSPEC command ``setplot area'' was used,
rather than plotting an unfolded spectrum, which serves to multiply by
the ratio residuals).  In this representation, it is especially clear
that the Fe XXV and Fe XXVI lines contain appreciable substructure,
hinted at in the first-order spectra.

Table 3--7 give the continuum and photoionization component parameters
for the fits made to each source, guided by the constraints detailed
above.  Errors (1$\sigma$ confidence levels) are given for the best-fit
model in each table.  Error assessment was extremely time
intensive, so errors are not given for inferior models.  However, the
$\chi^{2}$ fit statisic is quoted for every model, so that the
improvements obtained from adding components or complexity can be
assessed.

GRO J1655$-$40 can be taken as an example (see Figure 4 and Table 3).
Model 1655-3a achieves the best overall $\chi^{2}$ value
($\chi^{2}/\nu = 1135/1098 = 1.033$); it includes 3 photoionization
components, all blurred by Gaussian functions.  Model 1655-3b details
the effect of removing one photoionization component (one paired
absorption/emission zone), and fitting again.  The fit is clearly only
marginally worse without a third photoionization component.
Similarly, model 1655-3c removed the blurring functions that acted
upon the photoionized emission, and again only a marginally worse fit
is achieved.  From this, we can gather that only two components are
strongly required in the third-order spectrum of GRO J1655$-$40, and
that blurring is also not a strong requirement at this sensitivity.  However,
model 1655-3d removes another paired absorption/emission zone, and the
fit is significantly worse ($\chi^{2}/\nu = 1172/1109 = 1.058$).
Comparing 1655-3d to all of the others, it is clear that the second
zone latches onto a highly ionized, high-velocity component, with
$v/c = -0.0118(5)$.  In Figure 4, the Fe XXVI doublet is
evident at lower velocity, closer to its rest-frame energy; above
7.0~keV, another doublet split by ~20~eV is evident, and this is the
second, higher-velocity component found by the fit.

Models for the third-order spectrum of GRS~1915$+$105 achieve
qualitatively similar results (see Figure 5 and Table 4).  The overall
best-fit model includes three photoionization zones, each blurred with
a Gaussian function; however, the improvement over a model including
only two zones and no blurring is only marginal at this sensitivity.
At least as measured in the third-order spectrum of GRS~1915$+$105,
the disk wind is considerably slower than that launched in GRO
J1655$-$40.

The sensitivity in the third-order spectra of 4U~1630$-$472 (see
Figure 6 and Table 5) and H~1743$-$322 (see Figure 7 and Table 6) is
lower than that achieved in GRO~J1655$-$40 and GRS~1915$+$105.  In the
case of 4U 1630$-$742, two photoionization zones are only modestly
preferred over a single zone.  Two zones do not provide a significant
improvement over a single zone in fits to H 1743$-$322, though this is
the least sensitive spectrum in our sample.

For both GRO J1655$-$40 and GRS 1915$+$105, it is particularly
important to note the success of the improved, high-resolution XSTAR
grids.  In both cases, the Fe XXV intercombination and resonance lines
are correctly modeled in absorption (see Figures 4 and 5).  In GRO
J1655$-$40 in particular, the model correctly reproduces the 20~eV
spin-orbit splitting of the Fe XXVI line.  Similar structure is likely
present in the spectrum of GRS 1915$+$105 as well, as the Fe XXVI line
is asymmetric, but the splitting is likely partly obscured by velocity
components that overlap.  Evidence of complex
atomic structure in the third-order spectra of 4U 1630$-$472 and
H~1743$-$322 (see Figures 6 and 7) is less clear than in GRO
J1655$-$40 and GRS~1915$+$105, but hints of the Fe XXV intercombination
line in absorption line are evident in 4U 1630$-$472, and structure
may be evident in the Fe XXVI lines of both sources.

It must be noted that the ratio of Fe XXV intercombination and
resonance lines shows evidence of saturation, consistent with the gas
properties that emerge from the fits (see below).  For weak lines, the
intercombination line is only about 10\% as strong as the resonance
line, based on a ratio of their oscillator strengths.  The fact that
the intercombination line appears to be a few times stronger than this
expectation is simply the result of saturated line absorption,
providing a valuable constraint on the Fe XXV column density.

In summary, the third-order spectra have limited sensitivity; however,
they have verified the need for improved atomic data and higher
resolution grids.  The spectra establish a statistical basis for multiple
absorption zones (or, velocity components) even at the highest
possible spectral resolution.  There are even weak hints of
photoionized emission that might be broadened.  These are the general
lessons that we carry forward in subsequent modeling of the more
sensitive first-order spectra (see below).  In the case of GRO
J1655$-$40, the third-order spectrum makes the presence of a highly
ionized, high-velocity zone clear; this component explains
curvature on the blue side of the first-order Fe XXVI line profile
(see Figures 1--3), so we carry forward the velocity measured from the
third-order absorption spectrum in subsequent fits to the first-order
spectrum.

\subsection{Fits to the first-order spectra}

New fits to the first-order spectra of GRO J1655$-$40, GRS 1915$+$105,
4U 1630$-$472, and H~1743$-$322 are detailed in Tables 7--10, and the
best-fit model for each spectrum is shown in Figures 8--12.  The
best-fit model for each source is listed first in these tables, with
the suffix ``-1a''.  Models listed below the best-fit case explore the
effects of removing different model components in order to demonstrate
the statistical requirement for different levels of complexity.  Here again,
full errors are given for every component of the best-fit model for
each first-order spectrum, but errors were not calculated for inferior
models (though the resultant fit statistic is given).

Figures 8 and 9 may be compared directly in order to appreciate the
improvement achieved by allowing for multiple velocity components and
corresponding photoionized emission.  Figure 8 shows the first-order
spectrum of GRO J1655$-$40, fit with the best-fit model published in
Miller et al.\ (2008).  That model includes only one absorption zone,
and no photoionized emission.  Although it captures the general character
of the Fe K absorption lines, the data/model ratio clearly establishes
the need for a better approach.  In a statistical sense, the model is
also quite poor: $\chi^{2}/\nu = 1555/498 = 3.180$ (note that the
fitting band is not exactly the same as that considered in Miller et
al.\ 2008).  In strong contrast, the best-fit model from Table 7
(model 1655-1a) is shown in Figure 9; it achieves a vastly improved
fit: $\chi^{2}/\nu = 634/475 = 1.315$.  The latter model is far more
complex, but the statistical improvement (over
$19\sigma$) is large enough to justify this complexity.

This best-fit model for the first-order of GRO J1655$-$40 requires
three paired photoioinized absorption/emission zones, and blurred
emission.  The velocity of the most blue-shifted absorption component
is fixed to $v/c = 11.8 \times 10^{-3}$, based on fits to the
third-order spectrum.  The three components follow a sequence such
that column density and the absorption outflow velocity are
anti-correlated.  The photoionized emission components are not found
to be red-shifted, consistent with expectations, but broadening of
these components is strongly required (compare 1655-1a and 1655-1b; an
F-test indicates the improvement is significant at more than
$13\sigma$).  It is notable that the two slower components require the
maximum allowed broadening ($\sigma = 0.2$~keV), whereas the fastest
and most ionized component requires the least broadening ($\sigma =
0.03\pm 0.01$~keV).  Broadening is actually more important than the
addition of a third photoionization zone, in a statistical sense,
though the third component is nominally significant at more than the
$7.6\sigma$ level of confidence (compare 1655-1a and 1655-1d).

Fits to the first-order spectrum of GRS 1915$+$105 are detailed in
Table 8, and the best-fit model for the spectrum is shown in Figure
10.  Similar to 1655-1a, the best fit model -- 1915-1a -- requires
three photoionization absorption/emission pairings.  None of the
blue-shifted absorption components are found to flow as quickly as the
fastest component in GRO J1655$-$40.  Indeed, the component with the
highest column density has a small outflow velocity ($v/c =
0.2^{+0.1}_{-0.2} \times 10^{-3} = 60_{-60}^{+30}$~km/s).  The highest
velocity component is also the most ionized, again similar to GRO
J1655$-$40, and it is measured to have an outflow velocity of $v/c =
8.0\pm 0.2 \times 10^{-3} = 2400\pm 600$~km/s).  

In the first-order spectrum of GRS~1915$+$105, there is again evidence
that the photoionized emission pairs for the stronger absorption must
be broadened.  Model 1915-1b removes Gaussian smoothing functions from
the fit; comparing 1915-1b to 1915-1a, broadening is required at the
$11.2\sigma$ level of confidence via an F-test.  The requirement for
three zones, rather than just two, is also strong: model 1915-1d
considers only two zones, and 1915-1a is superior at more than the
5$\sigma$ level of confidence.

Fits to the first-order spectrum of 4U 1630$-$472 are detailed in
Table 9, and the best-fit model (1630-1a) is shown in Figure 11.  In
constrast to GRO J1655$-$40 and GRS 1915$+$105, the spectrum of 4U
1630$-$472 requires only two paired photoionization absorption and
emission components.  However, the same general trend is recovered:
the component with the highest column density has a lower blue-shift
and is less highly ionized, and while the second component has a lower
column density, it is more highly ionized and has a higher blue-shift.
These broad similarities betwen the black hole winds may signal common
properties and common launching radii and mechanisms.  The slower
absorption component in 4U 1630$-$472 has a velocity of $v/c = -0.9\pm
0.2 \times 10^{-3} = -270\pm 60$~km/s), and the second absorption
component has a velocity of $v/c = -7.0^{+3.0}_{-2.0} \times 10^{-3} =
-2100^{+900}_{-600}$~km/s).  The emission components paired to each of
these absorption components are not required to be red-shifted.

Comparing models 1630-1a and 1630-1b, broadening of the photoionized
emission components is required at the $5.7\sigma$ level of
confidence, via an F-test.  The requirement for photoionized emission
is also lower; comparing 1630-1c with 1630-1a shows that the addition
of emission components is only significant at the $3\sigma$ level of
confidence.  However, this likely under-predicts the actual
significance of these components, as broadening may be important to
evaluating the action of emission within the fit.  There are different
ways to evaluate the improvement achieved by considering two zones
rather than just one; depending on the specific models compared, the
improvement can be as low as $2\sigma$.  However, based on the broad
similarities in the properties of the components seen in GRO
J1655$-$40, GRS 1915$+$105, and 4U 1630$-$472, it is likely that the
emission components are real and stem from the same geometric
considerations.  Moreover, the observation of 4U 1630$-$472 did not
register the same high flux level as recorded in GRO J1655$-$40 and
GRS~1915$+$105; stacking observations of 4U 1630$-$472 woud likely
serve to boost the significance of the emission, but lies beyond the
scope of this paper.

The first-order spectrum of H~1743$-$322 is likely the simplest of
those considered, likely owing only to the fact that the column
density of the wind in this source is lower than in the other three.
In Figure 1, the Fe XXV and XXVI lines profiles appear to be
asymmetric.  A superposition of Fe XXV intercombination and resonance
lines from a single absorbing component would create a composite line
at first-order resolution that is asymmetric in the opposite sense
(stronger in a false ``blue'' wing).  The sense of the observed
asymmetry, and the explicit blue wing in the observed Fe XXVI line
profile signal multiple photoionization/velcity components
in the obseved disk wind.

Model 1743-1a in Table 10 includes two paired photoionization
components.  The fit achieved with this model is shown in Figure 12.
The lower velocity absorption component is outflowing at approximately
$300\pm 90$~km/s, the second component has a blue-shift of
$2100^{+900}_{-600}$~km/s.  The lower-velocity absorption component
carries a lower column density and a lower ionization parameter than
the high velocity component.  The high velocity, column density, and
ionization of the second component signal that it carries more mass
flux, and transmits more kinetic power.  The higher velocity component
is not strongly required via an F-test comparing 1743-1a and 1743-1d.
However, a fit to the Fe XXVI complex with two Gaussians suggests that
the higher-velocity component is signficant at the $3\sigma$ level of
confidence.  This spectrum is consistent with broadening of the
re-emission spectrum, but it is not statistically required at this low
sensitivity.

\subsection{Radius-focused Photoionization Modeling}

Some of the (re-) emission components found in the prior fits require
significant blurring, up to $\sigma \simeq 0.2$~keV.  At an energy of
6.7~keV, this translates to a speed of $0.03c$, or the Keplerian
orbital velocity at $r \simeq 1000~ {\rm GM/c}^{2}$.  Prior photoionization
modeling of GRO J1655$-$40 found similar launching radii, and indeed
similar radii are implied for several of the components reported in
this work.  However, it is important that all components be modeled as
carefully as possible in all aspects, in order to derive the most
accurate dynamical information.

At low resolution, and/or low sensitivity, Gaussian functions may be
sufficient to describe the emission line profile expected at $r \simeq
1000~ {\rm GM/c}^{2}$.  However, the line profile should actually be
asymmetric, primarily owing to Doppler shifts.  Gravitational
red-shifts may also be detectable, but this would require exquisite
data.  In order to better explore the ability of the data to reveal
the launching radius through Doppler broadening of its re-emitted
spectrum, we replaced the Gaussian functions in fits to the
first-order spectra with the ``rdblur'' function (Fabian et
al.\ 1989).

The ``rdblur'' model has important advantages over more recent models,
though it has lower resolution.  Models such as ``kerrconv''
(Brenneman \& Reynolds 2006) and ``relconv'' (Dauser et al.\ 2012)
only extend out to $r = 400~ {\rm GM/c}^{2}$ and $r = 1000~{\rm GM/c}^{2}$,
respectively.  In contrast, ``rdblur'' is analytic (meaning that it
runs quickly; it is based on ``diskline''; Fabian et al.\ 1989) and
can extend out to $r = 10^{5}~ {\rm GM/c}^{2}$, and beyond.  The ``rdblur''
model assumes a Schwarzschild black hole, but differences between the
Schwarzshild and Kerr metrics are negligible at the radii of interest.
For simplicity, we assumed a ``standard'' single power-law emissivity
profile of $J \propto r^{-3}$.  The local gas will emit isotropically,
in an $r^{-2}$ fashion.  We have modeled the absorption zones assuming
a constant density, which again argues for a $r^{-2}$ emissivity.  In
strict terms, though, the wind may have a $r^{-2}$ density profile,
requiring an overall $r^{-4}$ emissivity with radius.  However, much
of the absorption occurs in the inner portion of a gas with a falling
density profile.  Overall, $r^{-3}$ is a reasonable estimate,
intermediate between two reasonable possibilities.  This profile is
found in simulations of re-emission from a central source (Wilkins \&
Fabian 2012).  Last, a hard lower limit of $500 {\rm GM/c}^{2}$ was enforced in
the fits.

The results of fits to the first-order spectra with ``rdblur''
replacing Gaussian blurring functions are given in Table 11.  In a
statistical sense, the best fits with this model are comparable to
those with the Gaussian funtions (see Tables 7--10).  Models with
$-r1$ appended constrain the radius of all paired absorption/emission
zones to be the same.  These models could be considered a column
density-weighted average.  Models with $-r2$ appended allow each
photoionized emission zone to have an independent best-fit radius.
Only the spectra of GRO J1655$-$40 and GRS 1915$+$105 have the
sensitivity required to make such fits.  It is important to note that
Table 11 only gives the $1\sigma$ errors on every fit parameter,
including inner radii (in keeping with other errors in this work).
The range of radii covered by $3\sigma$ confidence limits is generally
a few times larger than the $1\sigma$ limits.

The radius-focused models in Table 11 are the most physical fits
presented in this work.  Figure 13 shows the relative importance of
absorption and emission in the spectra, and evidence of disk-like
P-Cygni profiles, based on the models in Table 11.  In Figure 13, the
photoionization components were removed from the total model for each
spectrum, leaving the ratio to the direct continuum emission.  The
total model including photoionization components is then plotted
through each ratio, to show the interplay of absorption and emission.

\subsection{Launching Radii and Outflow Properties}

Table 12 lists the estimated mass outflow rates and kinetic
luminosities for the best models in Table 11.  In all cases, the mass
outflow rate was calculated via $\dot{M}_{wind} = \Omega \mu m_{p} v L
/ \xi$, and the kinetic power was calculated via $L_{kin} = 0.5
\dot{M} v^{2}$ (where $\Omega$ and $L_{rad}$ are covering fraction and
radiated luminosity given in Table 2, the ionization parameter $\xi$
was measured using the XSTAR grids, $\mu$ is the mean atomic weight
and $\mu = 1.23$ was assumed, $m_{p}$ is the mass outflow rate, and
$v$ is the measured blue-shift of each component).  Eddington-scaled
quantities are tabulated assuming an accretion efficiency of $\eta =
0.1$ in the simple equation $L_{rad} = \eta \dot{M} c^{2}$.  In all
cases, a luminosity uncertainty of 50\% was included to estimate
outflow parameters, to account for uncertainties in source mass,
distance, and continuum model.

These calculations assumed a volume filling factor of unity, and in
that respect they represent a kind of upper limit.  The wind may be
clumpy, but there is no evidence of strong short-term variations that
would suggest clumping.  Indeed, the relative stability of the
absorption lines in GRO J1655$-$40 has even been utilized to constrain
the parameters of the binary system (Zhang et al.\ 2012; however, see
Madej et al.\ 2014).  Where variability has been observed in wind
aborption spectra, it appears to be more closely tied to changes in
the incident ionizing flux (e.g. Miller et al.\ 2006b) and/or changes
in large-scale geometry (Ueda et al.\ 2009, Neilsen et al.\ 2012a,b),
not tied to clumpiness.  In contrast, the clumpiness of the companion
wind in Cygnus X-1 is partly indicated by the distribution of strong
X-ray flux dips with orbital phase (e.g. Wen et al.\ 1999).  

The most ionized components in the wind do not necessarily carry the
greatest mass flux, but they do account for the greatest part of the
kinetic luminosity of the wind from each black hole.  The mass outflow
rates do not exceed the inferred mass inflow rate, but in some cases
the ratio approaches $\dot{M}_{wind}/\dot{M}_{accr} \simeq 0.3$
(e.g. zone 2 for 1915-r2 in Table 12).  In total, the total mass flux
and kinetic power carrried by the wind from each black hole generally
exceed prior estimates (e.g. King et al.\ 2013) by an order of
magnitude, or more.  This owes to the higher-velocity
components detected in our modeling of the spectra, since the mass
flux goes as $v$ and the kinetic power goes as $v^{2}$.  

On a componenent-by-component basis, Table 12 compares the radius
derived from photoionization modeling, $r_{phot} = \sqrt{(L/n\xi)}$,
to the radius derived from broadening of the emission line spectrum.
The radii derived from modeling the emission lines are generally
slightly smaller than those derived from photoionization modeling, but
the two measures generally agree to within a factor of a few.
Especially in view of the numerous uncertainties and issues related to
modeling the gas, this level of agreement is encouraging.  It is
consistent with a scenario wherein the wind retains much of the
rotation of the underlying disk, meaning that broadened (re-)emission
spectra can be used to estimate radii.  If this is correct, then the
winds are very close to the local escape speed.  

This is merely an initial attempt at such modeling, and some
inconsistencies remain.  Factoring in errors, the most substantial
discrepancy exists for ``zone 1'' in 1915-r2, where $r_{phot} =
17,000\pm 3000 {\rm GM/c}^{2}$, whereas $r_{blur} = 1200^{+600}_{-200}
{\rm GM/c}^{2}$.  This component is among the slowest detected in our sample
($v = -0.001c$), and it also has a fairly modest ionization (log $\xi$
= $4.10\pm 0.05$).  The photoionization radius of this component could
be brought into closer agreement with its dynamical radius by modeling
with a higher gas density.

None of the components described by Tables 11 and 12, nor indeed those
detailed in Tables 3--10, can be driven through radiation pressure.
The gas is simply too highly ionized for a force-multiplier effect
(e.g. Proga 2003).  Similarly, the optical depth to electron
scattering is too low for even the components with the highest column
densities to be driven in that manner.  This leaves only thermal
driving and magnetic forces as viable means of expelling the observed
winds.

Thermal winds can be driven from the Compton radius, given by $R_{C} =
(1.0\times 10^{10}) \times (M_{BH}/M_{\odot})/T_{C8}$ (where $T_{C8}$
is the Compton temperature in units of $10^{8}$~K), or perhaps from
$0.1~R_{C}$ (Begelman et al.\ 1983, Woods et al.\ 1996).
Approximating the Compton temperature by the disk blackbody color
temperature in Table 2, the smallest Compton radius inferred for any
source is $R_{C} \simeq 4.4\times 10^{11}~ {\rm cm}$.  This
corresponds to $4\times 10^{5}~ {\rm GM/c}^{2}$ for GRO J1655$-$40, or about
$3\times 10^{5}~ {\rm GM/c}^{2}$ for the other sources.  Even if a thermal
wind can be driven from $0.1 R_{C}$, this is still an order of
magnitude larger than the radii estimated in our spectral fits.
Theoretical work on thermal winds is ongoing, and recent work has
found that such winds may be denser and faster than previously
recognized (e.g. Higginbottom \& Proga 2015), but the results of our
analysis appear to confirm a role for magnetic driving.

The ratio of wind kinetic luminosity to radiated lumonsity is
also given in Table 12, in units of $10^{-3}$.  In this
metric, the outflow from GRO J1655$-$40 stands out for having tapped
into the accretion luminosity most efficiently.  Theoretical work by
Hopkins \& Elvis (2010) has found that AGN can effectively blow gas
from host galaxies, if just 0.5\% of the radiated luminosity couples
to drive the initial wind.  All of the outflows in our work are
inferred to have a kinetic luminosity below 0.5\% of the radiated
luminosity, but GRO J1655$-$40 is only a factor of a few below this
level, and the most ionized components in GRS 1915$+$105 and the
others are within an order of magnitude.  Since radiative force cannot
drive any of these flows, however, they may not serve as a guide to
the downstream power imparted to AGN winds, but rather only a guide to
the power available to initiate an AGN outflow that is later affected
by radiation.

\subsection{Wind emission from low-inclination disks}
Our results suggest that (re-)emission from equatorial disk winds is
important, even in sources viewed at high inclination, where
absorption lines are much stronger.  If this is correct, then emission
should be seen from the disk wind in sources that are viewed at low or
moderate angles, and any absorption should be very weak.  To test
this, we examined archival HETG spectra of GX 339$-$4 and XTE J1817$-$330 in
disk--dominated high/soft states.  Based on an absence of X-ray dips,
optical studies giving low K velocities and weak ellipsoidal
variations, and the detection of a one-sided jet in GX 339$-$4 (Gallo
et al.\ 2004), this source is likely viewed at relatively low
inclination angle.  Moderately complex X-ray absorption spectra have
previously been detected in GX 339$-$4; however, the observed lines do
not vary with the continuum, and it is likely that such lines originate
in the ISM (see, e.g., Juett et al.\ 2006, Pinto et al.\ 2013).

The prime difficulty with this test is that GX 339$-$4 and XTE
J1817$-$330 are {\it very} soft in the high/soft state, delivering
little sensitivity in the Fe K band compared to 4U 1630$-$472, GRO
J1655$-$40, H~1743$-$322, and GRS 1915$+$105.  The spectra appear to
generally lack the signal required to confirm or reject emission from
a disk wind.  The most promising evidence of features that may be
consistent with an equatorial wind seen in emission, is found in the
most sensitive observation of GX 339$-$4 (ObsID 4571; see Table 1).
Figure 14 shows the ratio of the data to a simple absorbed disk
blackbody plus power-law continuum in the 5--8~keV band (giving
$\chi^{2}/\nu = 353/337$; the data were binned to require 20 counts
per bin).  There is evidence of emission lines consistent with 6.7~keV
and 6.97~keV, corresponding to He-like Fe XXV and H-like Fe XXVI.

A significantly better fit to the spectrum of GX 339$-$4 is achieved
if the XSTAR emission component used in fits to GRO J1655$-$40 is
included in the model (see Figure 14).  The gas is consistent with
${\rm N}_{H} = 1.7\times 10^{23}~ {\rm cm}^{-2}$, log($\xi$) = 4.6,
and is blue-shifted by $v/c = 1700\pm 300$.  The fit is improved at
the 4.7$\sigma$ level of confidence, as determined by an $F$ test
($\chi^{2}/\nu = 323/334$).  The component normalization is measured to
be $K = 0.05\pm 0.01$, which suggests the wind lines are significant
at the $5\sigma$ level of confidence.

\section{Discussion and Conclusions}
We have re-analyzed sensitive high-resolution {\it Chandra} spectra of
accretion disk winds in four stellar-mass black holes.  A combination
of the resolution afforded by third-order spectra and improved
photoionization models enabled a better characterization of the lines,
and the detection of new atomic features.  The Fe XXVI Ly-$\alpha$
line, for instance, was revealed as a pair in GRO J1655$-$40, split by
spin-orbit coupling in the H-like atom.  Information from the
third-order spectra, and a more serious examination of the fitting
residuals left by single-zone models, revealed the need for 2--3 zones
in all cases.  Some of the additional wind zones reveal gas moving at
much higher speeds, leading to mass outflow rates and kinetic
luminosities that are much higher than previously appreciated.

Our analysis also finds evidence of (re-)emission from the
photoionized winds.  These emission spectra appear to be broadened by
the degree expected if the wind executes Keplerian orbital motion at
the photoionization radius.  Two independent lines of evidence then
point to small wind launching radii ($r_{wind} \simeq 10^{3}~
{\rm GM/c}^{2}$).  Such radii are inconsistent with thermal driving, and
imply a role for magnetic processes.  The radii, velocity, and
geometry of the winds we have analyzed bear strong similarities to the
BLR in AGN, and suggest a physical connection.  In this section, we
discuss these results and implications in a broader context, and
comment on how they can be tested in the future.
 
The inadequacy of a single absorption zone to fit the spectrum of GRO
J1655$-$40 has been noted previously.  Kallman et al.\ (2009) found
evidence of a highly ionized, high-velocity component in addition to a
slower, less-ionized component.  Neilsen \& Homan (2012) also found
evidence of two components, and suggested that multiple processes may
work in tandem to drive the disk wind in GRO J1655$-$40.  Our analysis
supports these results, but significantly extends them for GRO
J1655$-$40, and the other sources as well.  The prior work had not
considered re-emission from the wind.

Warm absorbers in Seyfert-1 AGN are typically modeled using 2--3
components with different ionization parameters and flow velocities
(e.g. Kaspi et al.\ 2002).  Our results strongly suggest that the same
procedure yields an excellent description of disk winds in
stellar-mass black holes.  This similarity may be partly superficial,
and a generic property of photoionized gas that spans at least a
modest range in radius.  Some recent simulations suggest that a
relatively smooth distribution of flow properties can give rise to
spectra that are typified by 2--3 distinct components (e.g. Giustini
\& Proga 2003; Higginbottom \& Proga 2015).  Better traces of wind
variations over time in future observations may be able to distinguish
between smooth wind parameters, and distinct physical zones.

Early observations of Seyfert-1 warm absorbers preferentially revealed
gas with low or moderate ionization, since both the intrinsic
Seyfert-1 spectra and the effective area of the HETGS peak at low
energy (see, e.g., Leet et al.\ 2001 concerning MCG-6-30-15).
However, as deep exposures were accumulated through subsequent
observing cycles, the reality and importance of more highly ionized
components became clear.  The strong Fe XXV and XXVI absorption lines
detected by Young et al.\ (2005) in MCG-6-30-15, for instance, reveal
a component that cannot be driven by radiation force, and which must
originate in a region consistent with the BLR.  In an analysis of NGC
4051, Krongold et al.\ (2007) also find evidence of an ionized wind
component that must originate in the BLR.  Recent work on NGC 4051 by
King et al.\ (2012) reaches a similar conclusion.  The detection of Fe
XXV and Fe XXVI absorption lines of comparable relative strengths in
spectra of NGC 3783 and NGC 3516 (Kaspi et al.\ 2002, Reeves et
al.\ 2004, Turner et al.\ 2008) likely signals that such components
are common in Seyfert warm absorbers, and generally consistent with
originating within the BLR or closer to the black hole.

Tremaine et al.\ (2014) recently examined the spectra of 20,000 AGN in
the Sloan Digital Sky Survey Data Release 7 quasar catalog.  In broad
terms, this work strongly suggests that the BLR is a disk wind,
potentially resolving the long-standing uncertainty regarding the
nature of the BLR.  Moreover, the specific results reported by Tremaine
et al.\ (2014) bear close similarities to the results that
have emerged from this and other recent work on stellar-mass black
hole disk winds:\\

\noindent$\bullet$ First, Tremaine et al.\ (2014) find that the broad H$\beta$ line has a
net gravitational redshift corresponding to Keplerian orbits at only
$10^{3-4}~ {\rm GM/c}^{2}$.  This range of BLR radii corresponds well
to the radii that result from our modeling of both the absorption and
(re-)emission spectra from stellar-mass black hole disk winds (see
Table 12).\\

\noindent$\bullet$ Second, Tremaine et al.\ (2014) confirm that the BLR is obscured at
high inclinations, suggesting that the BLR is equatorial.  Prior work
on disk winds in stellar-mass black holes strongly suggests that they
are equatorial (Miller et al.\ 2006a,b; King et al.\ 2012; Ponti et
al.\ 2012).  The P-Cygni profiles revealed in this analysis also require an equatorial geometry (see Figure 13).\\

\noindent$\bullet$ Third, Tremaine et al.\ (2014) note that the line of sight flow
velocities in the SDSS sample are far below local Keplerian
velocities.  The same is true of some absorption components in
stellar-mass black hole disk winds.  Our results make it clear that
this is a geometric effect, and that broadened emission components
confirm the small launching radii implied by photoionization modeling
of the wind absorption lines.  Simulations also suggest that geometry
and viewing angle can lead to disparities between observed speeds and
full gas speeds (Giustini \& Proga 2012).\\

\noindent {\em In essence, equatorial X-ray disk winds may be the BLR
  for stellar-mass black holes.}\\

If (re-)emission from winds in stellar-mass black holes is really
analogous to the BLR in AGN, then it should be seen as the dominant
signature of the disk wind in sources viewed at lower inclinations.
The necessary sensitivity is not commonly available in archival
observations of low-inclination stellar-mass black holes, but the
expected wind emission signature is detected at approximatly the
$5\sigma$ level of confidence in the best {\it Chandra}/HETG spectrum
of GX 339$-$4 in the high/soft state (see Figure 14).  Prior fits to
the {\it Chandra}/HETG spectrum of MAXI J1305$-$704 also required
re-emission (Miller et al.\ 2014), and this may also support the
picture and connections suggested by the present analysis.

It is possible that evidence of (re-)emission from stellar-mass black
hole disk winds has previously been identified as disk reflection.
Ueda et al.\ (2009) required {\it distant} reflection to fit the {\it
  Chandra} and {\it RXTE} spectra of GRS 1915$+$105 considered in this
work.  In the fits made by Ueda et al.\ (2009), the narrowness of the
Fe K emission line(s) required the inner edge of the reflecting gas to
extend no closer $r_{in} \geq 400~{\rm GM/c}^{2}$.  This is a strange outcome
given that the disk is expected to extend to the ISCO in soft states
close to the Eddington limit (e.g. Esin et al.\ 1997).  Indeed, Ueda
et al.\ (2009) report inner disk radii of 71--120~km, or
4.6--7.9~GM/c$^{2}$ (for $M_{BH} = 10.1~M_{\odot}$; Steeghs et al.\ 2013).
The column densities observed in the wind from GRS 1915$+$105 are
high, and perhaps not entirely different than the columns present in
the atmosphere of an ionized accretion disk where reflection is
expected.  

The detection of multiple components in each spectrum -- some with
much higher outflow speeds than identified before -- has served to
increase the mass outflow rate and kinetic power for the winds in each
systems over prior estimates (see Table 12; also see King et
al.\ 2013).  Apart from the small launching radii now implied by two
aspects of our modeling, the larger outflow rates also point to the
need for magnetic processes -- not just thermal driving -- to expel the
winds (Miller et al.\ 2008; also see Woods 1996).  The kinetic power
in the winds does not represent a large fraction of the radiative
Eddington limit, but the kinetic power of the wind in GRO J1655$-$40
represents the highest fraction of the observed radative luminosity,
$L_{W}/L_{rad} \simeq 0.002$.  In this metric, it still stands out
relative to the other winds in the sample.

If disk winds in stellar-mass black holes connect to the BLR and the
most ionized components of Seyfert-1 warm absorbers, our results have
consequences for AGN.  For instance, broadened re-emission from winds
with particularly high column densities, like the highly ionized
component in MCG-6-30-15 (Young et al.\ 2005), should also be
detected.  With the imminent launch of {\it Astro-H}, more ionized
outflows should be detected in Seyfert-1 AGN.  If those flows are like
stellar-mass black hole disk winds, the total outflow rate and kinetic
power should increase by orders of magnitude over estimates based on
low-ionization outflows detected in early {\it Chandra} and {\it
  XMM-Newton} exposures.  AGN outflows will then come closer to
supplying the feedback needed to halt star formation in host bulges.
Current estimates based on the detections of low-ionization wind
components find that warm absorbers may be unable to do supply the
necessary feedback (see, e.g., Blustin et al.\ 2005).

The high-velocity components that we have discovered ($v/c \simeq
0.01$, and just below) begin to bridge the gap between $few \times
100~{\rm km}~ {\rm s}^{-1}$ outflows, and ``ultra-fast outflows'' (or,
UFOs; see Tombesi et al.\ 2010).  Only one UFO has been detected in a high
resolution gratings spectrum: components found in the stellar-mass
black hole IGR J17091$-$3624 have velocities of $v/c = 0.03$ and $v/c =
0.05$ (King et al.\ 2012).  The UFOs found in CCD spectra of AGN are
generally of low statistical significance, and it is less certain what
future observations of putative UFOs will reveal.  The $v/c \simeq
0.08$ outflow claimed in the {\it XMM-Newton} spectrum of PG
1211$+$143 was not recovered in a more sensitive observation with {\it
  NuSTAR} (Zoghbi et al.\ 2015); however, prior evidence of a strong
outflow in PDS 456 may be confirmed with {\it NuSTAR} (Nardini et
al.\ 2015).  If UFOs originate even closer to the central engine than
the components at the center of this work -- if they are launched from
$few \times 100~ {\rm GM/c}^{2}$ -- then the central engine may not be taken
as a point source, and rotation of the wind may imprinted on the {\it
  absorption} lines in {\it Astro-H} spectra.

Going forward, several things can be done to improve upon our approach,
and to test our results:

We have only undertaken a cursory analysis of the extent to which
reflection could account for the emission-line flux that appears to be
dominated by the wind.  Fits to the first-order spectra of GRO
J1655$-$40 and GRS 1915$+$105 with the {\it relxill} reflection model
(Dauser et al.\ 2013) replacing wind re-emission produce dramatically
worse fits.  Similarly, fits to RXTE spectra taken simultaneously with
these {\it Chandra} observations show reflection-like residuals, but
fits with wind absorption and re-emission replacing putative
reflection produce vastly improved fits.  More work is needed in this
regard.  It is possible that reflection of very steep power-law
spectra from a highly ionized disk could account for some of the flux
that our models currently ascribe to re-emission from the wind, but
new reflection models are required to test this.

Very deep observations with {\it Chandra} that permit more sensitive
third-order HETG spectra can help to test our results.  Deep
observations of GRS 1915$+$105 and other transients in very soft
states can help to verify or falsify the picture that has emerged.
Deep observations of transients viewed at low inclination, such as
GX~339$-$4, are also important.  Spectra dominated by narrow emission
lines in soft states would confirm the picture that we have developed;
tight limits on expected line features would help to reject the same
picture.

{\it Astro-H} should be able to undertake the best disk wind and
reflection spectroscopy, for both stellar-mass black holes and
Seyfert-1 AGN (see, e.g., Miller et al.\ 2014, Kaastra et al.\ 2014).
The resolution and sensitivity afforded by the {\it Astro-H}
calorimeter may be able to detect red-shifts in X-ray wind emission
lines (as per the gravitational red-shifts found in the BLR; Tremaine
et al.\ 2014), if the winds are not outflowing at very high speeds.
Whereas dispersive spectrometers offer progressively lower resolution
with increasing energy, a calorimeter offers progressively higher
resolution with increasing energy.  Thus, {\it Astro-H} should be very
sensitive to the fastest, most ionized wind components that carry the
most mass and kinetic power.  The broad bandpass of {\it Astro-H},
will afford {\it NuSTAR}-like sensitivity up to 50--100~keV, enabling
contributions from disk reflection to be measured simultaneously.
These capabilities should enable definitive tests of our results.

\vspace{0.1in}

JMM is grateful to Michael Nowak and Keith Arnaud for helpful
discussions and assistance with XSPEC and ISIS.  JMM acknowledges
helpful discussions with Kayhan Gultekin, Dipankar Maitra, and Mark
Reynolds.  We thank the anonymous referee for comments that improved
this work.

\clearpage

\begin{figure}
\hspace{0.5in}
\includegraphics[scale=0.8,]{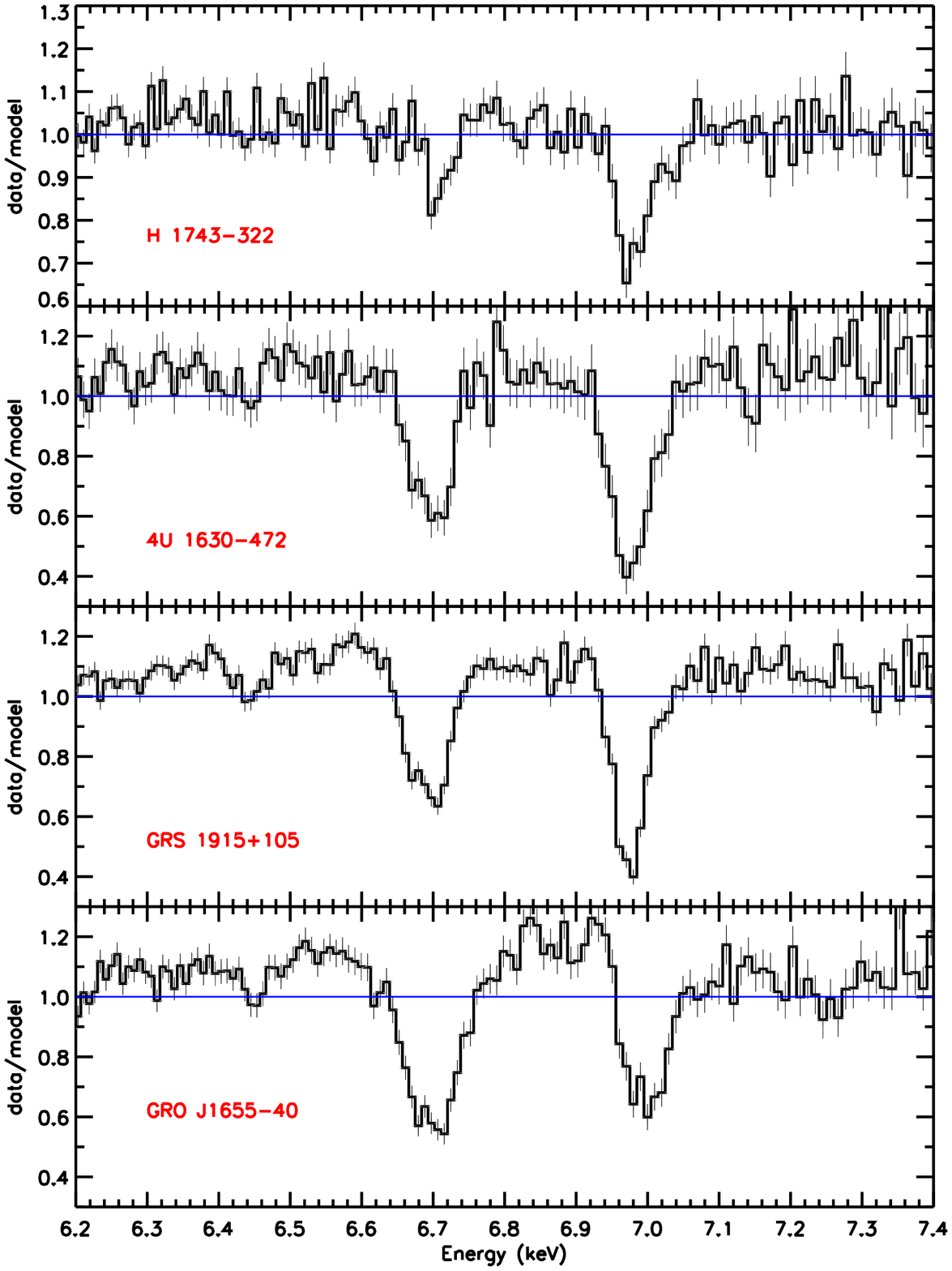}
\figcaption[t]{\footnotesize The first-order spectra of H 1743$-$322,
  4U 1630$-$472, GRS 1915$+$105, and GRO J1655$-$40 are each shown as
  a ratio to the best-fit continuum model, initially ignoring the Fe K
  band.  Each model included disk blackbody and power-law components,
  with power-law indices constrained via broad-band fits to
  simultaneous RXTE data, where possible.  The He-like Fe XXV and
  H-like Fe XXVI absorption lines in each spectrum show non-Gaussian
  structure, indicating contributions from related lines and/or
  multiple velocity components.  Evidence of weak emission redward
  of the absorption lines is also apparent, especially in GRO
  J1655$-$40 and GRS 1915$+$105, and suggestive of disk-like P-Cygni
  profiles.}
\end{figure}
\medskip

\clearpage

\begin{figure}
\hspace{0.5in} \includegraphics[scale=0.8,]{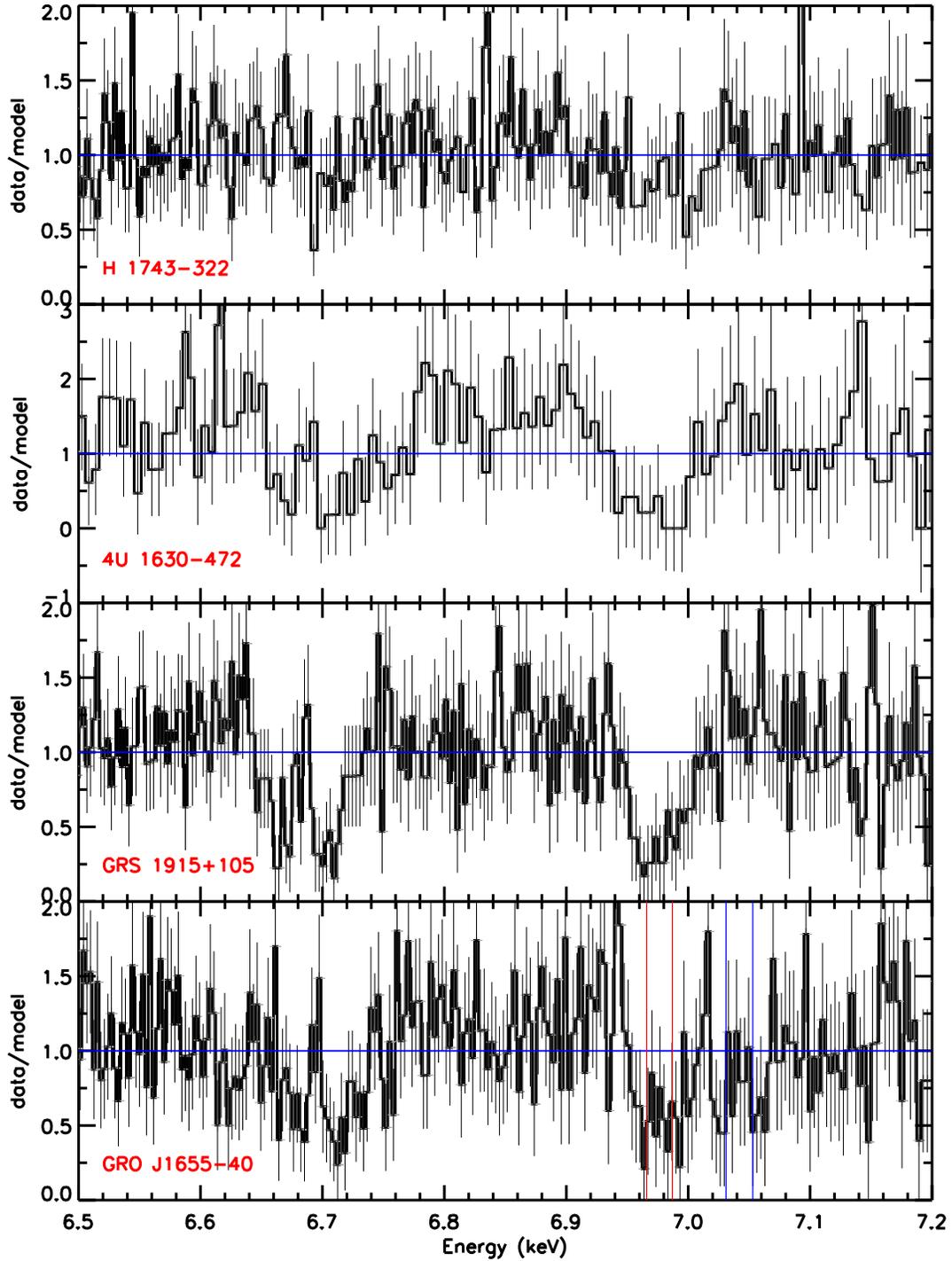}
\figcaption[t]{\footnotesize The third-order spectra of H 1743$-$322,
  4U 1630$-$472, GRS 1915$+$105, and GRO J1655$-$40 are each shown as
  a ratio to a simple continuum, as per the prior figure.  Third-order
  spectra have three times higher resolution than first-order spectra;
  the resolution of these spectra is only a factor of a few lower than
  the spectra anticipated from {\it Astro-H}.  In the most sensitive
  spectra, the Fe XXV complex is resolved into distinct lines, with
  the intercombination line seen in absorption (in addition to the
  resonance line).  Especially in GRO J1655$-$40, but also evident in
  other spectra, the Fe XXVI line is resolved as a doublet - for the
  first time - owing spin-orbit coupling (analytical theory predicts a
  separation of just 0.02~keV, as observed). The doublet is evident
  just below 7.0~keV and just above 7.0~keV in GRO J1655$-$40, clearly
  indicating a high velocity component.  Vertical lines indicate the theoretical separation for the pairs.}
\end{figure}
\medskip

\clearpage

\begin{figure}
\hspace{0.5in} \includegraphics[scale=0.8,]{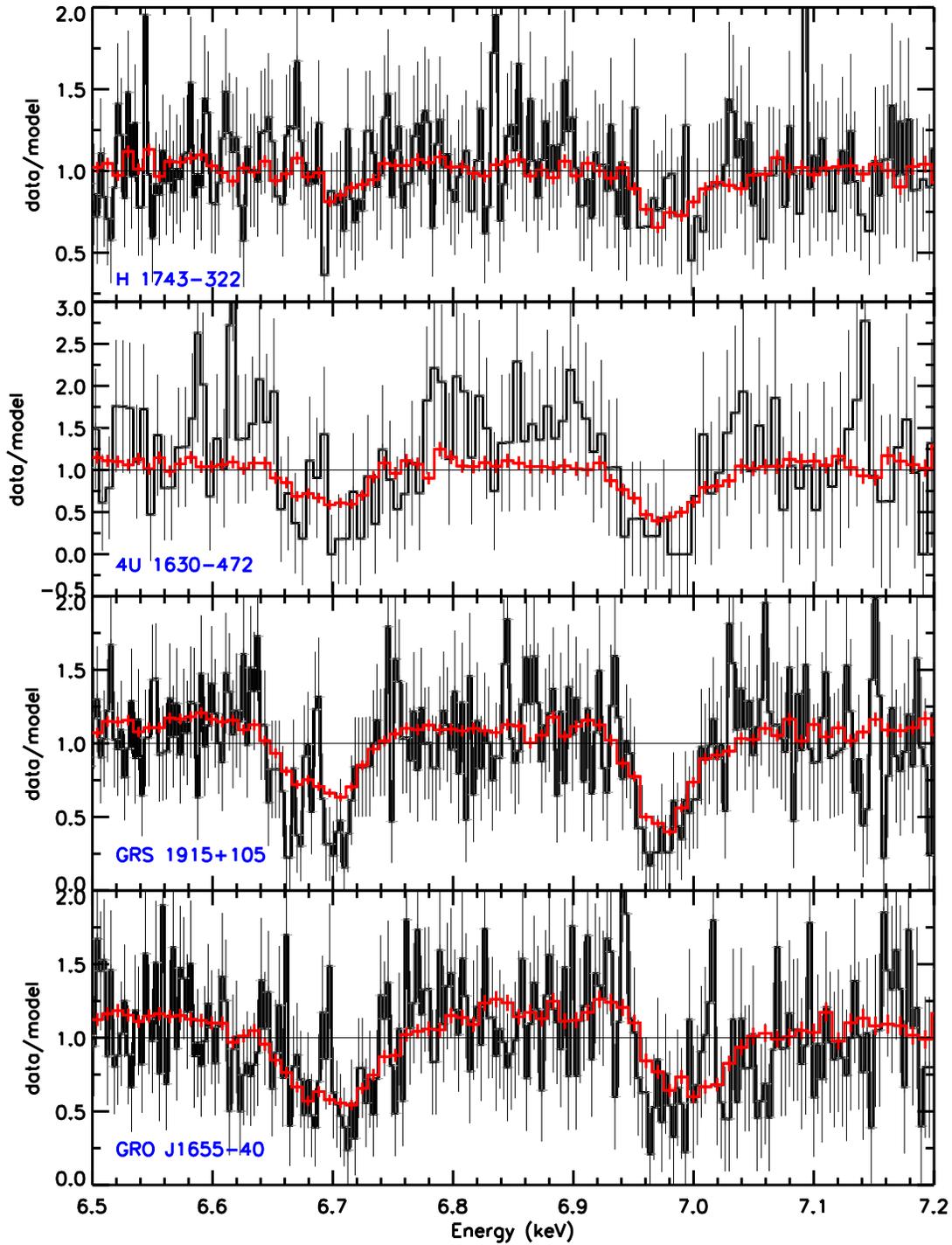}
\figcaption[t]{\footnotesize The first-order data/model ratios of
  H~1743$-$4322, 4U 1630$-$472, GRS 1915$+$105, and GRO J1655$-$40
  (shown in red) are plotted on top of the third-order data/model
  ratios (shown in black).  The individual lines seen in the
  third-order spectra clearly map onto the structure and asymmetries
  seen in the more senstiive first-order spectra.}
\end{figure}
\medskip

\clearpage

\begin{figure}
\includegraphics[scale=0.6,angle=-90]{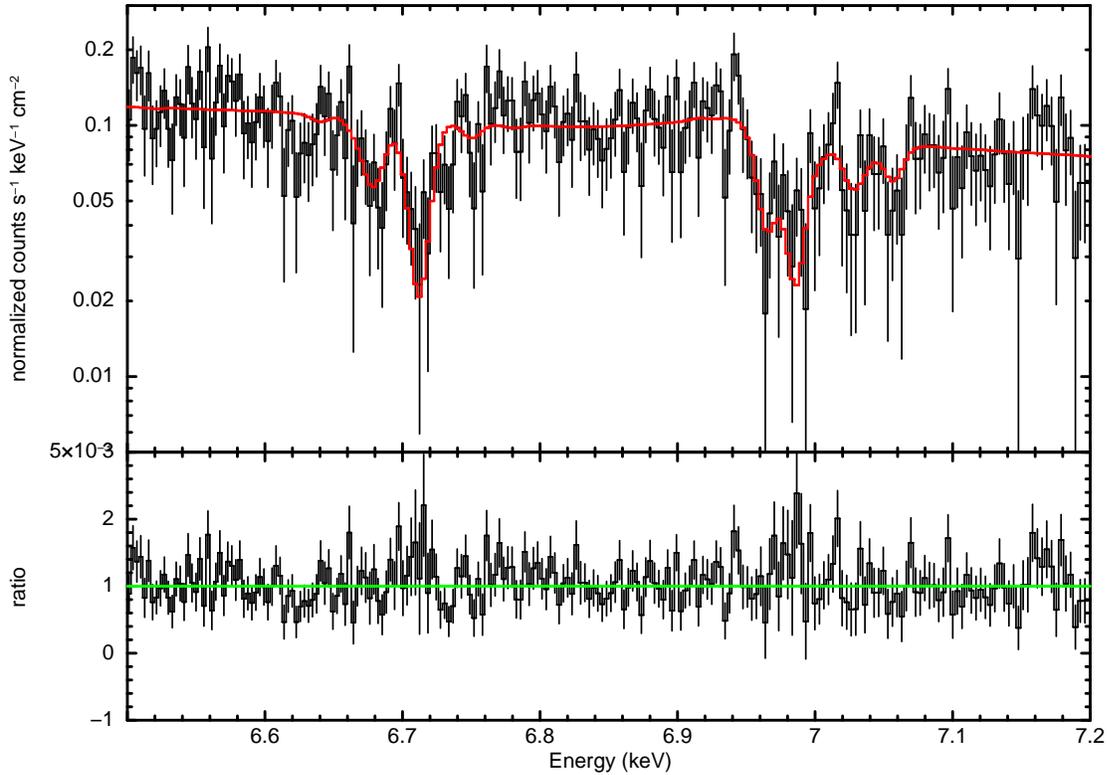}
\figcaption[t]{\footnotesize The third-order spectrum of GRO
  J1655$-$40, fit with a three-zone model (1655-3a; see
  Table 3).  The Fe XXV complex (6.67-6.70 keV) is clearly resolved,
  and the second-strongest line is the intercombination line.  The Fe
  XXVI lines are individually resolved, and independent pairs at
  smaller and larger shifts are evident.  The pair at the highest blue
  shift has an unexpected flux ratio but this is partly accounted for
  by the inclusion of a third absorption zone; a third zone is also
  required in fits to the lower-resolution but more sensitive
  first-order spectrum.}
\end{figure}
\medskip

\begin{figure}
\includegraphics[scale=0.6,angle=-90]{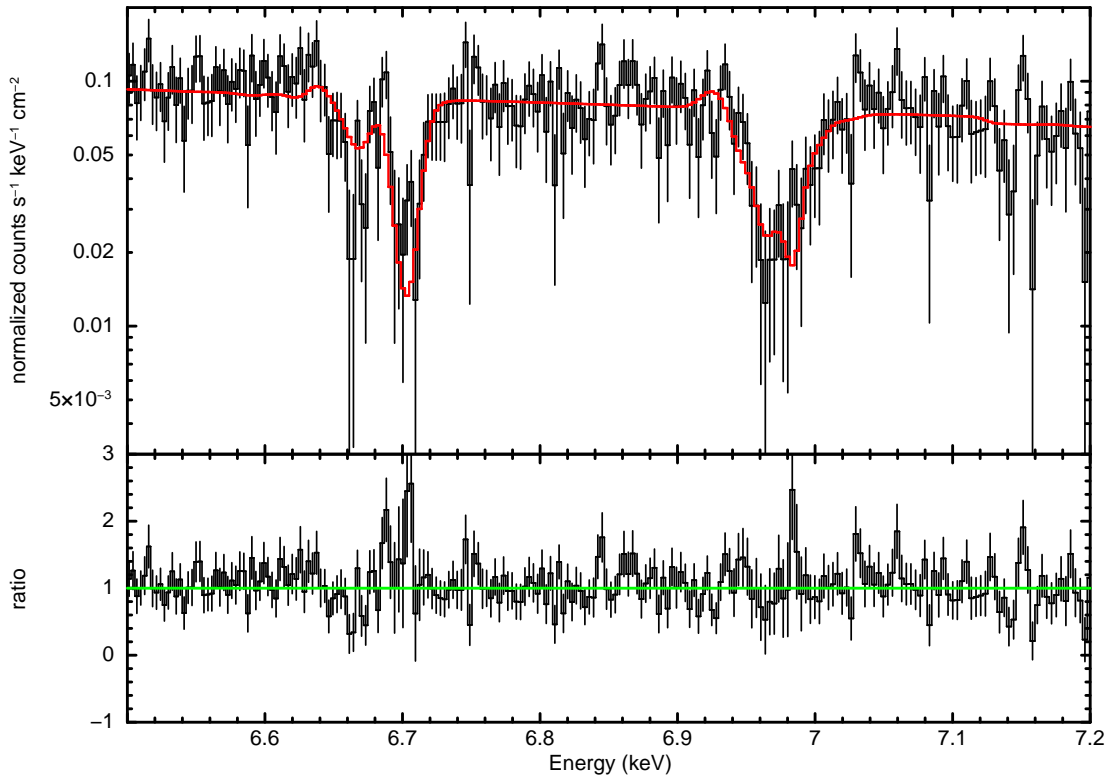}
\figcaption[t]{\footnotesize The third-order spectrum of GRS
  J1915$+$105, fit with a three-zone model (1915-3a; see
  Table 4).  The Fe XXV complex (6.67-6.70 keV) is clearly resolved,
  and the second-strongest line is the intercombination line.  The Fe
  XXVI line shows structure, including a tail to the blue indicating a
  high velocity component.}
\end{figure}
\medskip

\clearpage

\begin{figure}
\includegraphics[scale=0.6,angle=-90]{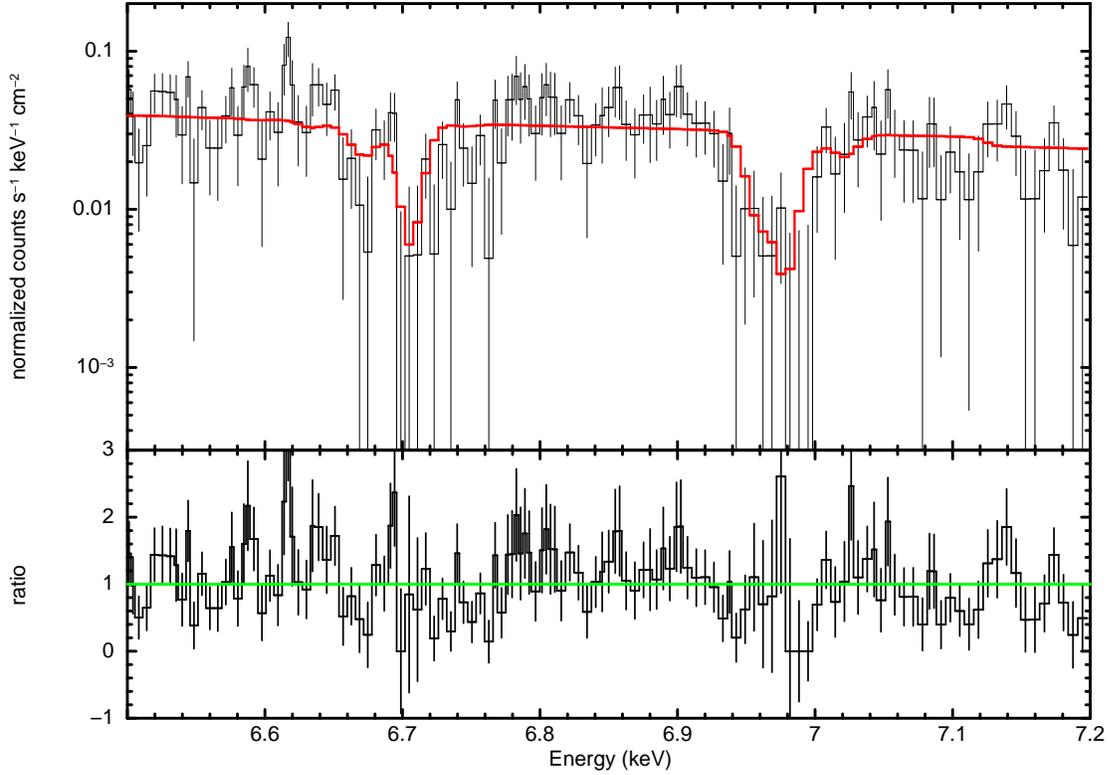}
\figcaption[t]{\footnotesize The third-order spectrum of
  4U~1630$-$472, fit with a two-zone model (1630-3a; see
  Table 5).  The spectrum was rebinned for visual clarity, in the plot
  above.  The sensitivity of this spectrum is considerably less than
  that seen in GRO J1655$-$40 or GRS 1915$+$105, but this spectrum
  clearly indicates the need for two components in fits to the more
  sensitive first-order spectrum of 4U 1630$-$472.}
\end{figure}
\medskip

\begin{figure}
\includegraphics[scale=0.6,angle=-90]{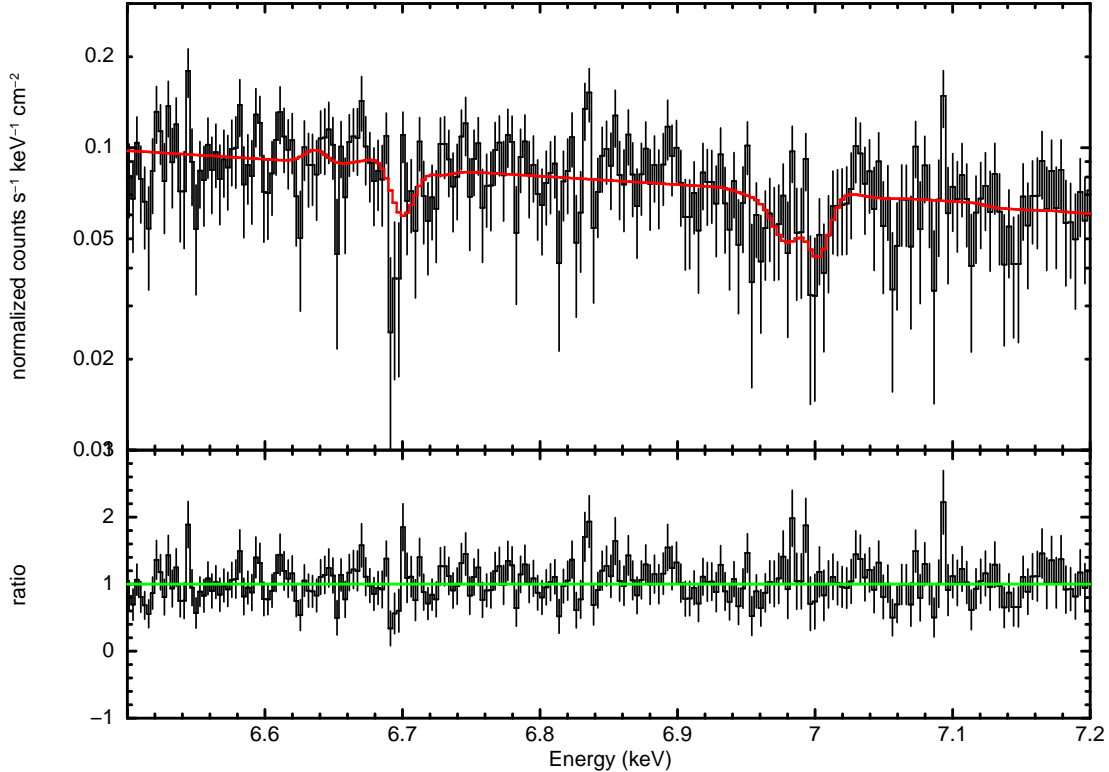}
\figcaption[t]{\footnotesize The third-order spectrum of H~1743$-$322,
  fit with a two-zone model (1743-3a; see Table 6).  The
  sensitivity of this spectrum is considerably less than that seen in
  GRO J1655$-$40 or GRS 1915$+$105, but the large blue-shift in the Fe
  XXVI line and asymmetry in the Fe XXV complex are apparent. This
  spectrum clearly indicates the need for two components in fits to
  the more sensitive first-order spectrum of H~1743$-$322.}
\end{figure}
\medskip

\clearpage

\begin{figure}
\includegraphics[scale=0.6,angle=-90]{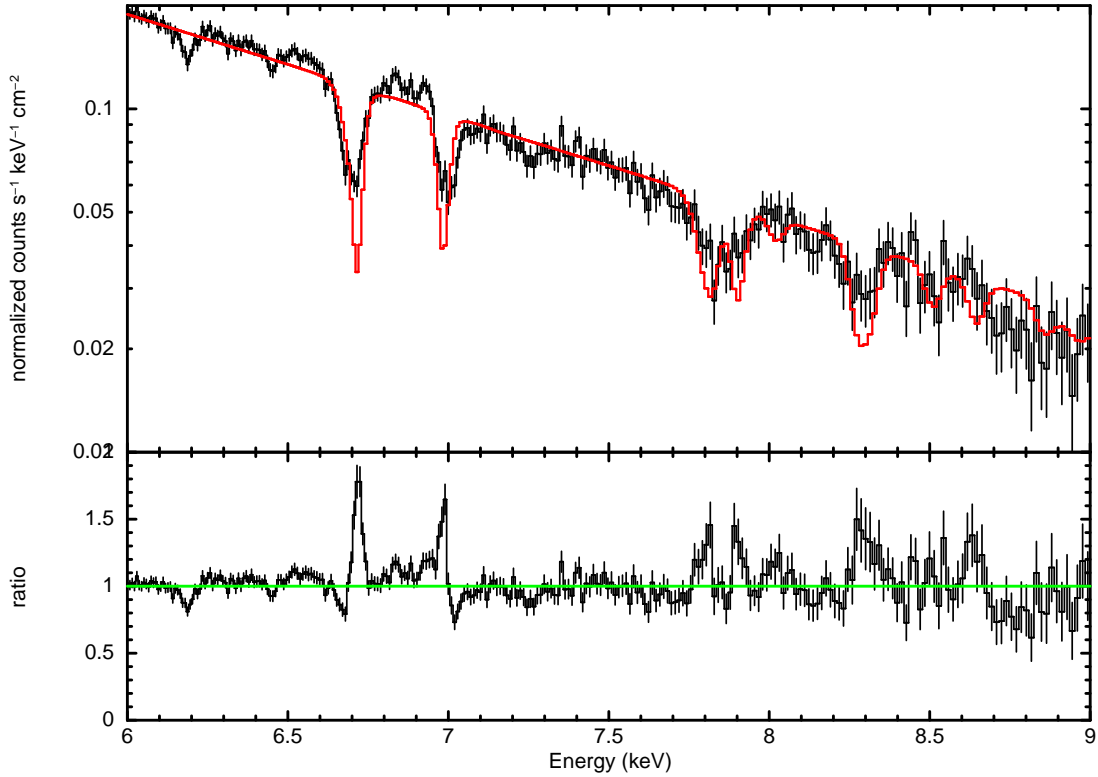}
\figcaption[t]{\footnotesize The first-order spectrum of GRO
  J1655$-$40 is shown here, fit with the best model for the Fe K band
  in Miller et al.\ (2008).  Emission components are not included in
  the model, which is comprised of a single absorption zone.  This
  model was vastly superior to others considered in Miller et
  al.\ (2008), though it was not an excellent description of the data.
  Using the procedures in this paper, the overall fit achieved is
  markedly worse ($\chi^{2}/\nu = 1555/489 = 3.180$) than the
  multiple-zone models including corresponding emission, detailed in
  Table 7 (model 1655-1a gives $\chi^{2}/\nu = 624/475 = 1.315$).}
\end{figure}
\medskip

\begin{figure}
\includegraphics[scale=0.6,angle=-90]{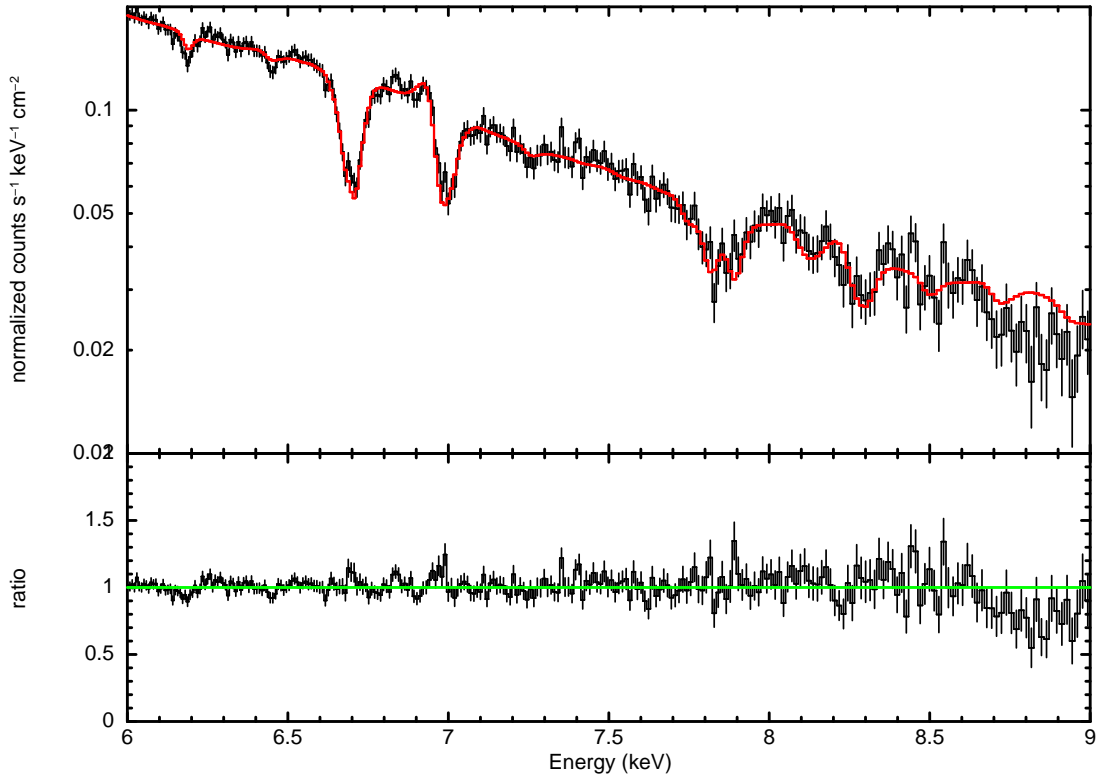}
\figcaption[t]{\footnotesize The first-order spectrum of GRO
  J1655$-$40 is shown here, fit with model 1655-1a in Table 7.  The
  model includes three absorption zones, and corresponding emission
  lines from gas with the same properties.  Blurring of the emission
  components by factors commensurate with Keplerian rotation at the
  photoionization radius is required by the data.  The fit is vastly
  superior to the simple model published in Miller et al.\ (2008) and
  shown in the prior figure.  }
\end{figure}
\medskip

\clearpage

\begin{figure}
\includegraphics[scale=0.6,angle=-90]{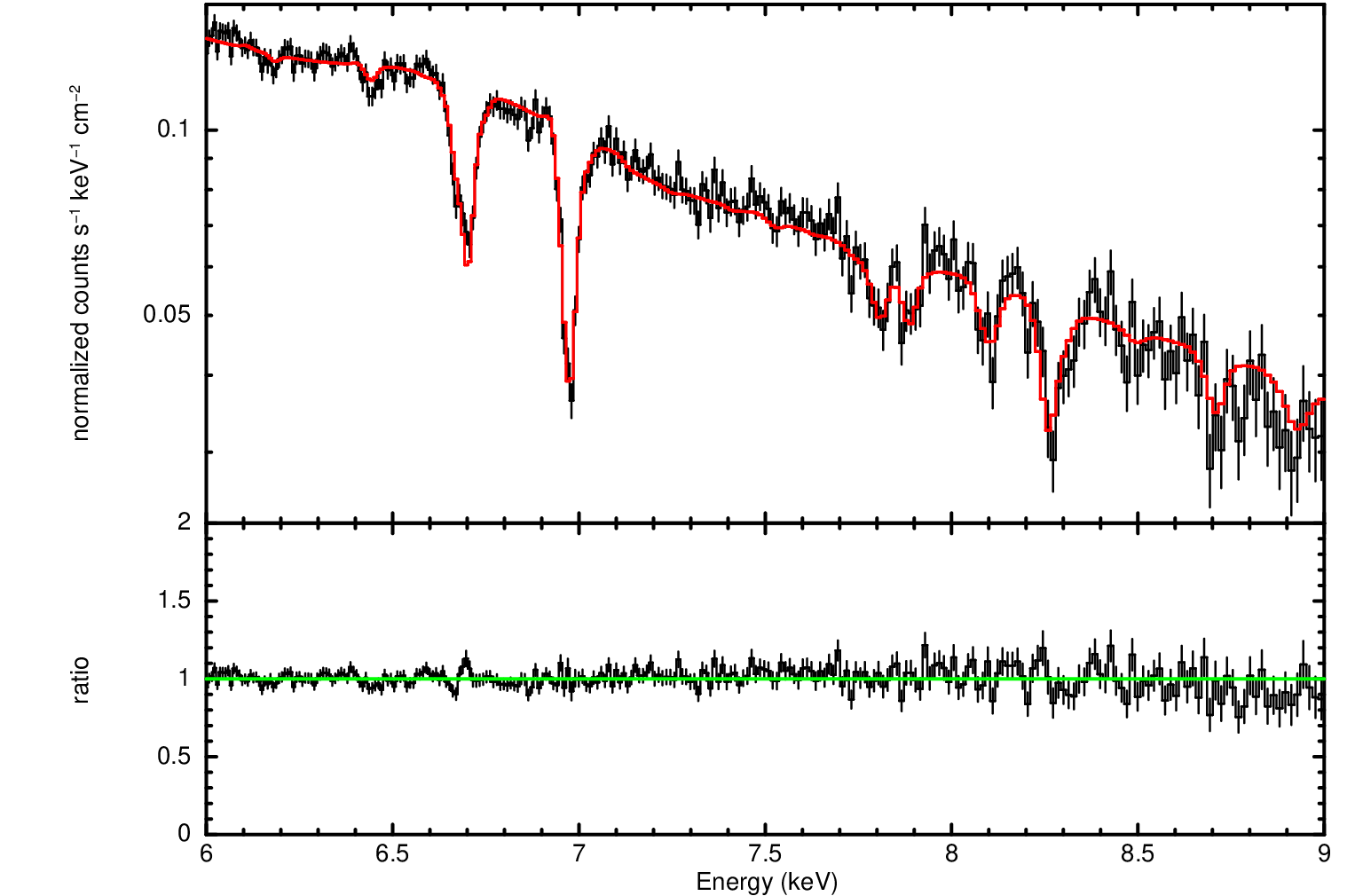}
\figcaption[t]{\footnotesize The first-order spectrum of GRS~1915$+$105
  is shown here, fit with model 1915-1a in Table 8.  The model
  includes three absorption zones, and corresponding emission lines
  from gas with the same properties.  Blurring of the emission
  components by factors commensurate with Keplerian rotation at the
  photoionization radius is required by the data.}
\end{figure}
\medskip

\clearpage

\begin{figure}
\includegraphics[scale=0.6,angle=-90]{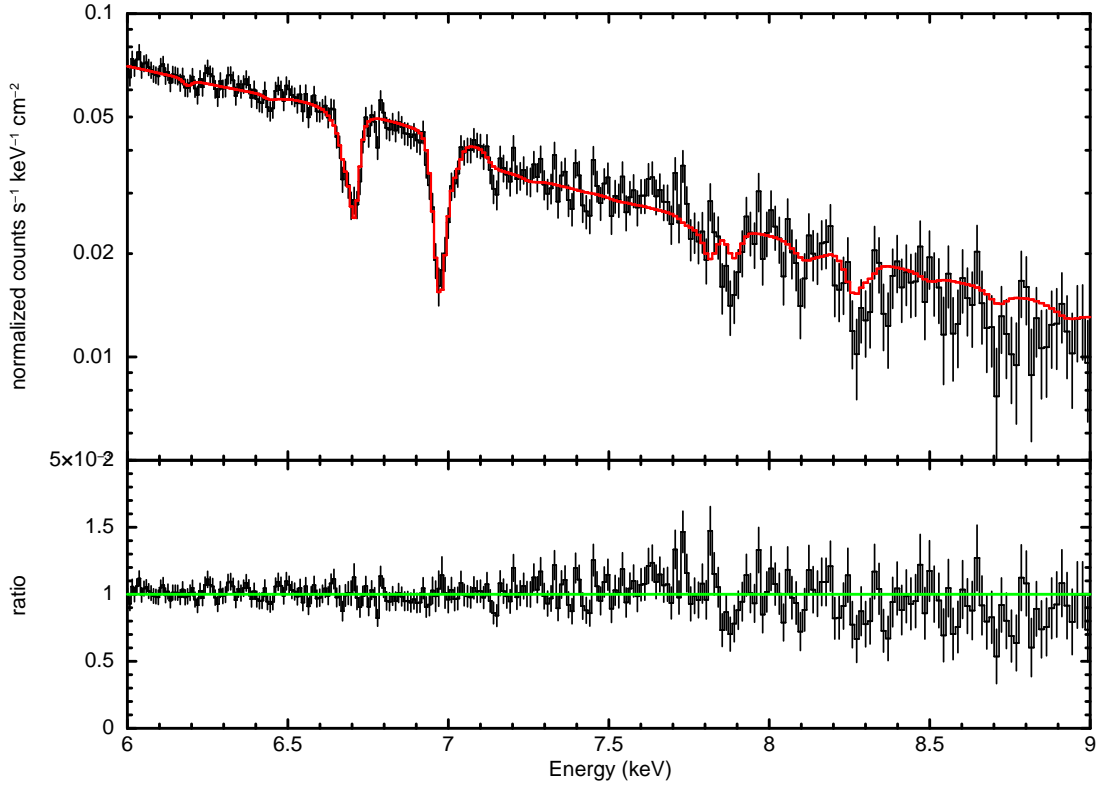}
\figcaption[t]{\footnotesize The first-order spectrum of 4U~1630$-$472
  is shown here, fit with model 1630-1a in Table 9.  The model
  includes three absorption zones, and corresponding emission lines
  from gas with the same properties.  Blurring of the emission
  components by factors commensurate with Keplerian rotation at the
  photoionization radius is required by the data.}
\end{figure}
\medskip

\begin{figure}
\includegraphics[scale=0.6,angle=-90]{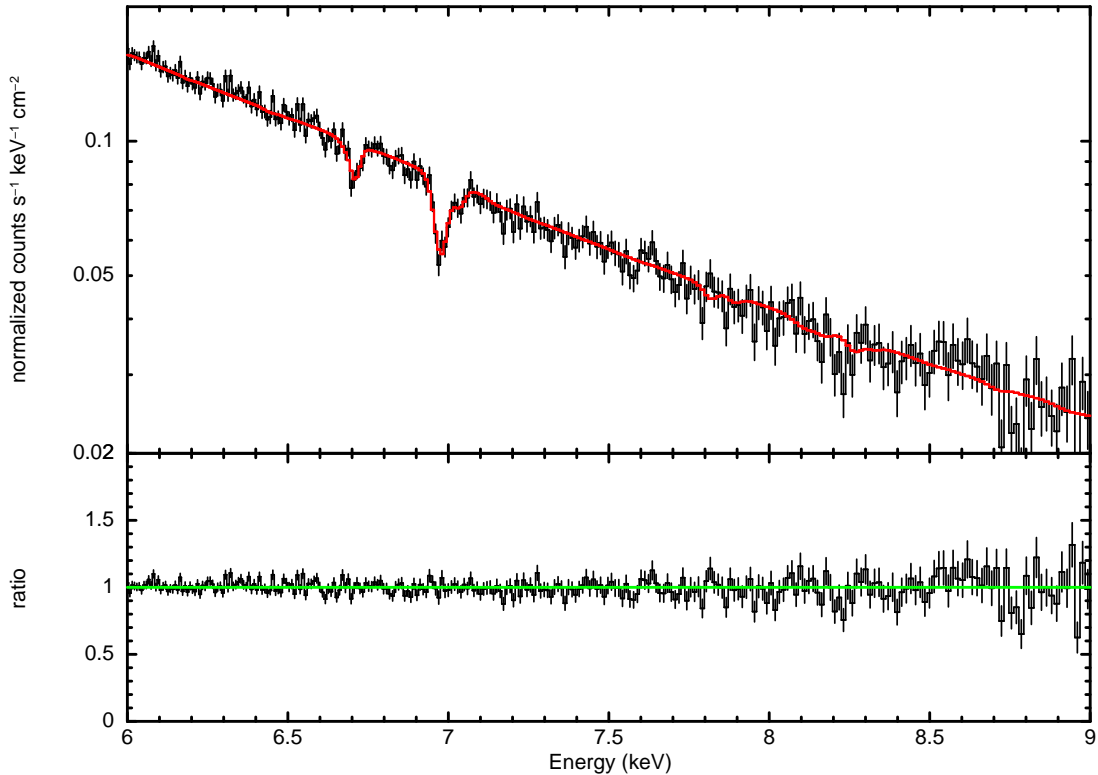}
\figcaption[t]{\footnotesize The first-order spectrum of H~1743$-$322
  is shown here, fit with model 1743-1a in Table 10.  The model
  includes two absorption zones, and corresponding emission lines from
  gas with the same properties.  The two-component model is able to
  fit asymmetries in the absorption lines that cannot be described
  with a single zone.}
\end{figure}
\medskip

\clearpage

\begin{figure}
\hspace{0.5in}
\includegraphics[scale=0.8,]{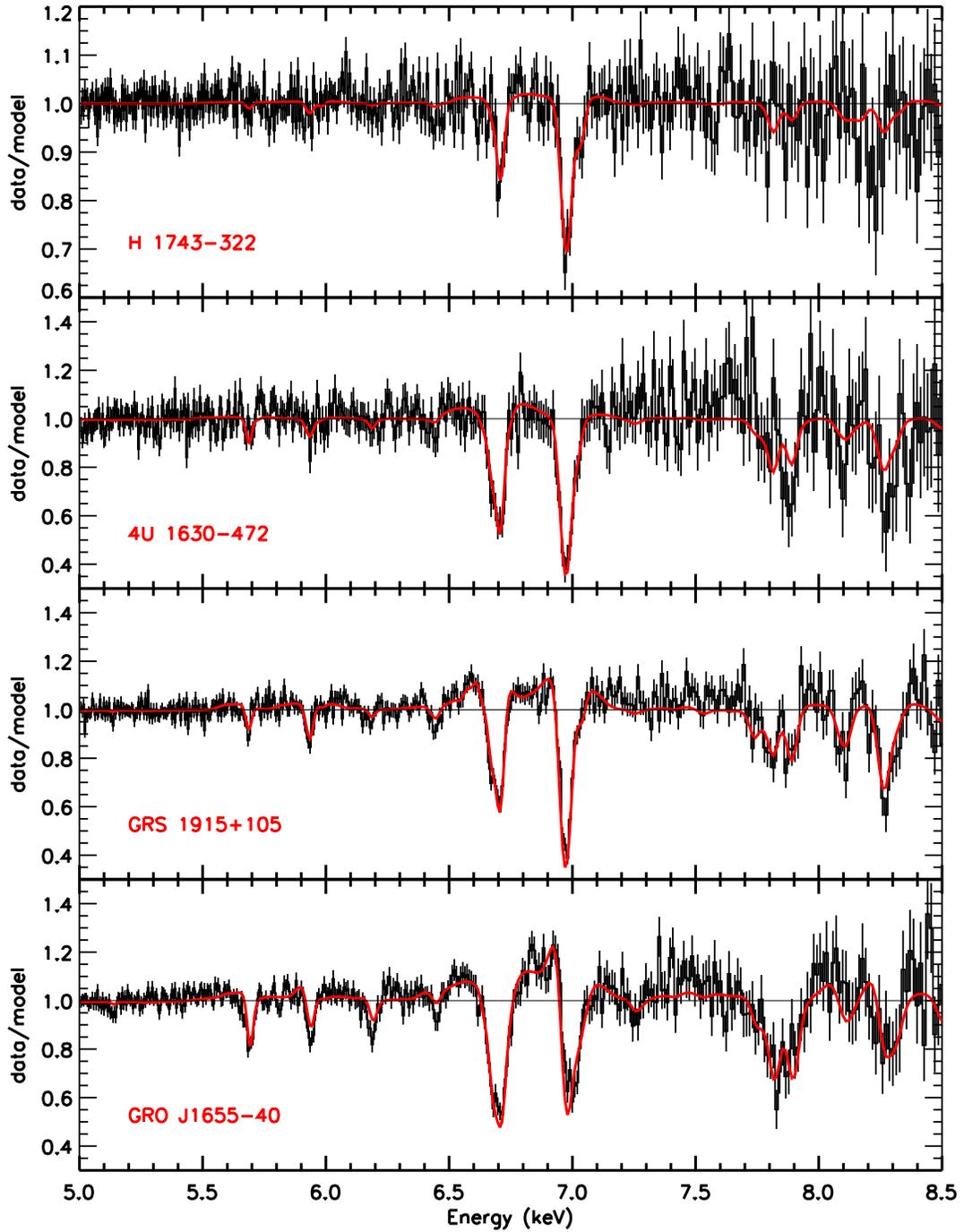}
\figcaption[t]{\footnotesize The first-order spectra of H 1743$-$322,
  4U 1630$-$472, GRS 1915$+$105, and GRO J1655$-$40 are each shown as
  a ratio to the best-fit continuum models from Table 11.  To
  construct each ratio, the photoionized absorption and emission
  components were removed from the total spectral model.  The model
  with the photoionized components is then plotted through the ratio,
  to illustrate the relative importance of absorption and emission in
  the spectra.  Evidence of disk-like P-Cygni profiles is strong in
  GRO J1655$-$40 and GRS 1915$+$105, and tentative in 4U 1630$-$472
  and H~1743$-$322.}
\end{figure}
\medskip

\clearpage

\begin{figure}
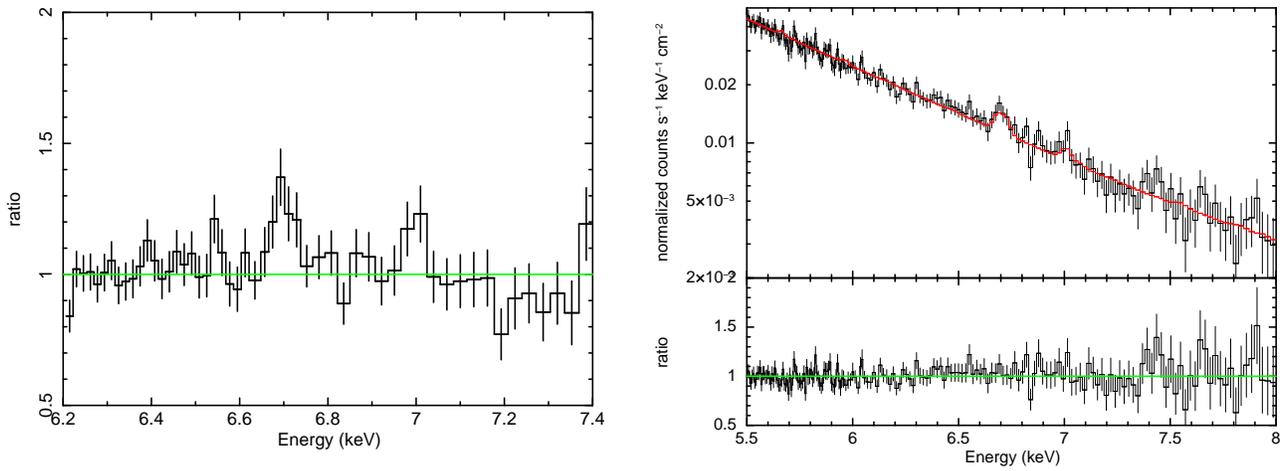

\includegraphics[scale=0.33,angle=-90]{f14a.ps}
\includegraphics[scale=0.33,angle=-90]{f14b.ps}
\figcaption[t]{\footnotesize A sensitive {\it Chandra}/HETG
  observation of GX~339$-$4 in the high/soft state may show evidence
  of a disk wind seen in emission (see Section 4.9).  LEFT: The
  data/model ratio from a simple disk blackbody plus power-law fit.
  The ratio shown in this figure contains modest evidence of He-like
  Fe XXV and H-like Fe XXVI emission lines (6.70~keV and 6.97~keV,
  respectively).  RIGHT: The spectrum of GX~339$-$4, fit with a model
  including the XSTAR wind emission component used in fits to GRO
  J1655$-$40 (see Tables 3, 7, 11).  The inclusion of the wind
  component is significant at the $5\sigma$ level of confidence.  The
  spectrum requires the emission component to be blue-shifted by $v =
  1700\pm 300$~km/s, consistent with a disk wind.}
\end{figure}

\clearpage

\begin{table}[t]
\caption{Sources and Observation Characteristics}
\begin{footnotesize}
\begin{center}
\begin{tabular}{lllll}
Source &   Observation     &   Duration &   Start Date &  Selected References \\    
   ~   &       ~           &   ($10^{3}$~s) &  (YYYY-MM-DD)        & ~  \\
\tableline
4U 1630$-$472   &  13715   &  29.3          & 2012-01-20 &  Neilsen et al.\ 2014 \\
GRO J1655$-$40  &  5461    &  65.6          & 2005-04-01 & Miller et al.\ 2006, 2008; Kallman et al.\ 2009 \\ 
H 1743$-$322    &  3803    &  46.5          & 2003-05-01 & Miller et al.\ 2006, 2012 \\
GRS 1915$+$105  & 7485     &  47.4          & 2007-08-14 & Ueda et al.\ 2009, Neilsen \& Lee 2009 \\ 
\tableline
GX 339$-$4      & 4571     & 35.3           & 2004-10-28 & -- \\
\tableline
\end{tabular}
\vspace*{\baselineskip}~\\ \end{center} 
\tablecomments{The table above lists basic parameters for each of the
  four main observations considered in this work, followed by a single
  observation of GX 339$-$4 that was searched for wind emission
  features.  The columns list the source name, {\it Chandra} ObsID,
  total raw exposure duration, start time, and up to three papers that
  have also considered the given observation.}
\vspace{-1.0\baselineskip}
\end{footnotesize}
\end{table}
\medskip

\begin{table}[t]
\caption{XSTAR Photoionization Grid Input Parameters}
\begin{footnotesize}
\begin{center}
\begin{tabular}{lllllll}
Source &   Luminosity    &    kT &   log(n) &  $v_{turb}$  & ${\rm A}_{Fe}$ & $f = \Omega/4\pi$\\     
   ~   & ($10^{38}~ {\rm erg}~ {\rm s}^{-1}$) & (keV) & ~  &  ${\rm km}~ {\rm s}^{-1}$ & ~ & ~ \\
\tableline
4U 1630$-$472   &   1.1 &        1.5    &    14.0 &    400  &         1.0 & 0.5 \\
GRO J1655$-$40  &   0.5 &        1.3    &    14.0 &    300  &        1.0 & 0.2 \\
H 1743$-$322    &  2.0  &       1.2     &   14.0 &     400  &         1.0 & 0.5 \\
GRS 1915$+$105  &  8.0  &       1.9     &   14.0 &     400  &        2.0 & 0.5 \\
\tableline
\end{tabular}
\vspace*{\baselineskip}~\\ \end{center} 
\tablecomments{The table above lists the critical input parameters
  used to create grids of high-resolution XSTAR photoionziation
  models.  In each case, a blackbody spectral input form was adopted,
  with temperatures based on continuum fits to each spectrum with a
  disk blackbody model.  A gas density of $10^{~14}~ {\rm cm}^{-3}$
  was assumed for all sources, based on the one case in this sample
  wherein such a direct measurement was possible (GRO J1655$-$40, see
  Miller et al.\ 2008).  An elevated Fe abundance, ${\rm A}_{Fe}$, was
  assumed for GRS 1915$+$105 based on prior work on HETG spectra (Lee
  et al.\ 2002). See the text for details the covering factor $f$.}
\vspace{-1.0\baselineskip}
\end{footnotesize}
\end{table}
\medskip

\clearpage

\begin{table}[t]
\caption{Fits to the Third Order Spectrum of GRO J1655$-$40}
\begin{footnotesize}
\begin{center}
\begin{tabular}{lllllll}
\tableline
Model & Notes & Zone 1 & Zone 2 & Zone 3 & Continuum & $\chi^{2}/\nu$ \\
\tableline
Example & 3 zones, emis. incl., blurring & ${\rm N}_{1}$ & ${\rm N}_{2}$ & ${\rm N}_{3}$ & kT (keV) & $500/500=1$ \\
 ~      &         ~         &  ${\rm log}(\xi_{1})$    & ${\rm log}(\xi_{2})$     & ${\rm log}(\xi_{3})$     & ${\rm K}_{disk}$ &   ~       \\
 ~      &         ~         &  $v_{abs, 1}/c$  & $v_{abs, 2}/c$   & $v_{abs, 3}/c$  &  $\Gamma$       &    ~       \\
 ~      &         ~         &  $v_{emis, 1}/c$ & $v_{emis, 2}/c$  & $v_{emis, 3}/c$ &   ${\rm K}_{pow}$ &     ~      \\
 ~      &         ~         &  $\sigma_{emis, 1}$ & $\sigma_{emis, 2}$ & $\sigma_{emis, 2}$      & ~  &    ~  \\
 ~      &         ~         &  ${\rm K}_{emis, 1}$ &  ${\rm K}_{emis, 2}$ &  ${\rm K}_{emis, 3}$ & ~  &  ~    \\

\tableline
1655-3a      & 3 zones, emis. incl., blurring & $22(4)$ & $5^{+15}_{-2}$ & $16^{+2}_{-4}$ & 1.3* & $1135/1098 = 1.033$ \\
 ~      &         ~         &  $4.42(5)$    & $4.7_{-0.2}^{+1.3}$     & $5.1(1)$     & 400 &   ~       \\
 ~      &         ~         &  $-1.8(1)$  & $-7.5(5)$   & $-11.8(5)$  &  3.5*       &    ~       \\
 ~      &         ~         &  $1.2$ & $3.4$  & $2.0$ &   22 &     ~      \\
 ~      &         ~         &  $0.19$ & $3^{+4}_{-3}$ & $0^{+12}$      & ~  &    ~  \\
 ~      &         ~         &  $0.1$ &  $0.2_{-0.1}^{+3}$ &  $5(3)$ & ~  &  ~    \\
\tableline
1655-3b      & 2 zones, emis. incl., blurring & $46(5)$ & $16(5)$ & -- & 1.3* & $1139/1101 = 1.033$ \\
 ~      &         ~         &  $4.8(1)$    & $5.5(2)$     & --    & 400(50) &   ~       \\
 ~      &         ~         &  $-1.8(1)$  & $-11.4(5)$   & --  &  3.5*       &    ~       \\
 ~      &         ~         &  $0^{+1.8}$ & $0^{+11}$  & -- &  84(3) &     ~      \\
 ~      &         ~         &  $0.2_{-0.1}$ & $0.03(1)$ & --      & ~  &    ~  \\
 ~      &         ~         &  $0.15(5)$ &  $5(2)$ & -- & ~  &  ~    \\
\tableline
1655-3c      & 2 zones, emis. incl., no blurring & $46$ & $7.5$ & -- & 1.3* & $1144/1103 = 1.197$ \\
 ~      &         ~         &  $4.8$    & $4.9$     & --    & 309 &   ~       \\
 ~      &         ~         &  $-1.6$  & $-11.5$   & --  &  3.5*       &    ~       \\
 ~      &         ~         &  $0$ & $0$  & -- &  37 &     ~      \\
 ~      &         ~         &  -- & -- & --      & ~  &    ~  \\
 ~      &         ~         &  0.1* & 2.7 & -- & ~  &  ~    \\
\tableline
1655-3d      & 1 zone, emis. incl., no blurring & $13$ & $--$ & -- & 1.3* & $1172/1109 = 1.058$ \\
 ~      &         ~         &  $4.4$    & --     & --    & 340 &   ~       \\
 ~      &         ~         &  $-1.9$  & --   & --  &  3.5*       &    ~       \\
 ~      &         ~         &  $0$ & --  & -- &  87 &     ~      \\
 ~      &         ~         &  -- & -- & --      & ~  &    ~  \\
 ~      &         ~         &  0.1* & -- & -- & ~  &  ~    \\
\tableline
\end{tabular}
\vspace*{\baselineskip}~\\ \end{center} 
\tablecomments{The table above details fits made to the third-order
  spectrum of GRO~J1655$-$40 in the 5.0--8.0~keV band, using a
  high-resolution XSTAR photoionization grid.  The ``example'' model
  explains the parameter values listed for subsequent models.  Where
  ``--'' appears, the component was not included in the model.
  Asterisks indicate that the parameter was frozen to the indicated
  value in the model.  Errors are quoted only for the best model(s).
  The hydrogen equivalent column density, ${\rm N}$, is given in units
  of $10^{22}~ {\rm cm}^{-2}$ for each absorption/emission zone.  The
  ionization parameter ($\xi = L/nr^{2}$) for the emission/absorption
  zone is quoted in log format for clarity.  When emission components
  are included in a model, the values of ${\rm N}$ and $\xi$ for each
  zone were fixed to same parameters in the absorption component.
  Negative velocity shifts indicate blue-shifts; positive velocity
  indicate red-shifts.  Velocities are listed in units of $10^{-3}$.
  Values of $\sigma$ in a given model indicate the width of Gaussian
  blurring of an emission component in units of keV, using the
  ``gsmooth'' model in XSPEC.  All instances of ``K'' indicate the
  flux normalization of a given component.  Please see the text for
  additional details.}
\vspace{-1.0\baselineskip}
\end{footnotesize}
\end{table}
\medskip

\clearpage

\begin{table}[t]
\caption{Fits to the Third Order Spectrum of GRS 1915$+$105}
\begin{footnotesize}
\begin{center}
\begin{tabular}{lllllll}
\tableline
Model & Notes & Zone 1 & Zone 2 & Zone 3 & Continuum & $\chi^{2}/\nu$ \\
\tableline
Example & 3 zones, emis. incl., blurring & ${\rm N}_{1}$ & ${\rm N}_{2}$ & ${\rm N}_{3}$ & kT (keV) & $500/500=1$ \\
 ~      &         ~         &  ${\rm log}(\xi_{1})$    & ${\rm log}(\xi_{2})$     & ${\rm log}(\xi_{3})$     & ${\rm K}_{disk}$ &   ~       \\
 ~      &         ~         &  $v_{abs, 1}/c$  & $v_{abs, 2}/c$   & $v_{abs, 3}/c$  &  $\Gamma$       &    ~       \\
 ~      &         ~         &  $v_{emis, 1}/c$ & $v_{emis, 2}/c$  & $v_{emis, 3}/c$ &   ${\rm K}_{pow}$ &     ~      \\
 ~      &         ~         &  $\sigma_{emis, 1}$ & $\sigma_{emis, 2}$ & $\sigma_{emis, 2}$      & ~  &    ~  \\
 ~      &         ~         &  ${\rm K}_{emis, 1}$ &  ${\rm K}_{emis, 2}$ &  ${\rm K}_{emis, 3}$ & ~  &  ~    \\

\tableline
1915-3a      & 3 zones, emis. incl., blurring & $40_{-10}^{+20}$ & $20_{-10}^{+10}$ & $3_{-2}^{+12}$ & 1.9(1) & $1059/1094 = 0.967$ \\
 ~      &         ~         &  $3.95(5)$    & $6.0_{-0.5}$     & $4.27(5)$     & 53(1)x &   ~       \\
~      &         ~         &  $-0.3_{-0.6}^{+0.3}$  & $-0.3^{+0.3}$   & $-4(1)$  &  3.0*       &    ~       \\
 ~      &         ~         &  $0.0^{+0.3}$ & $4(2)$  & $0.0^{+0.2}$ &  $0^{+4}$ &     ~      \\ 
 ~      &         ~         &  $0.2_{-0.1}$ & $0.04(4)$ & $0.00^{+0.03}$      & ~  &    ~  \\
 ~      &         ~         &  $0.10^{+0.03}$ &  $0.1^{+0.2}$ &  $0.4(2)$ & ~  &  ~    \\
\tableline
1915-3b      & 2 zones, emis. incl., blurring & $50_{-20}^{+10}$ & $5_{-2}^{+5}$ & -- & 1.9(1) & $1067/1101 = 0.969$ \\
 ~      &         ~         &  $3.94(7)$    & $4.4(2)$     & --    & 53(1) &   ~       \\
 ~      &         ~         &  $-0.4_{-0.1}^{+0.4}$  & $-4(1)$   & --  &  3.0*       &    ~       \\
  ~      &         ~         &  $0^{+2}$ & $0.0^{+0.6}$  & -- & $0^{+4}$  &     ~      \\
 ~      &         ~         &  $0.2_{-0.05}$ & $0.1^{+0.4}$ & --      & ~  &    ~  \\
 ~      &         ~         &  $0.13(3)$ &  $0.7(3)$ & -- & ~  &  ~    \\
\tableline
1915-3c      & 2 zones, emis incl., no blurring & $46$ & $7.5$ & -- & 1.9(1) & $1111/1103 = 1.197$ \\
 ~      &         ~         &  $4.8$    & $4.68$     & --    & 53 &   ~       \\
 ~      &         ~         &  $-1.6$  & $-3.3$   & --  &  3.0*       &    ~       \\
 ~      &         ~         &  $0$ & $0$  & -- &  0.0 &     ~      \\
 ~      &         ~         &  -- & -- & --      & ~  &    ~  \\
 ~      &         ~         &  0.1 & 0.1 & -- & ~  &  ~    \\
\tableline
1915-3d      & 2 zones, no emis., no blurring & $6.4$ & $12.5$ & -- & 1.9 & $1093/1107 = 0.988$ \\
 ~      &         ~         &  $3.89$    & $4.55$     & --    & 54 &   ~       \\
 ~      &         ~         &  $-0.4$  & $-2.9$   & --  &  3.0*       &    ~       \\
 ~      &         ~         &  -- & --  & -- &  0 &     ~      \\
 ~      &         ~         &  -- & -- & --      & ~  &    ~  \\
 ~      &         ~         &  -- & -- & -- & ~  &  ~    \\

\tableline
1915-3e      & 1 zone, no emis., no blurring & $10.6$ & $--$ & -- & 1.9 & $1101/1110 = 0.992$ \\
 ~      &         ~         &  $3.9$    & --     & --    & 54 &   ~       \\
 ~      &         ~         &  $-0.7$  & --   & --  &  3.0*       &    ~       \\
 ~      &         ~         &  -- & --  & -- & 0 &     ~      \\
 ~      &         ~         &  -- & -- & --      & ~  &    ~  \\
 ~      &         ~         &  -- & -- & -- & ~  &  ~    \\
\tableline
\end{tabular}
\vspace*{\baselineskip}~\\ \end{center} 
\tablecomments{The table above details fits made to the third-order
  spectrum of GRS~1915$+$105 in the 5.0--8.0~keV band, using a
  high-resolution XSTAR photoionization grid.  The ``example'' model explains the parameter values
  listed for subsequent models.  Where ``--'' appears, the component
  was not included in the model.  Asterisks indicate that the
  parameter was frozen to the indicated value in the model.  Errors
  are quoted only for the best model(s).  The hydrogen equivalent
  column density, ${\rm N}$, is given in units of $10^{22}~ {\rm
    cm}^{-2}$ for each absorption/emission zone.  The ionization
  parameter ($\xi = L/nr^{2}$) for the emission/absorption zone is
  quoted in log format for clarity.  When emission components are
  included in a model, the values of ${\rm N}$ and $\xi$ for each zone
  were fixed to same parameters in the absorption component.  Negative
  velocity shifts indicate blue-shifts; positive velocity indicate
  red-shifts.  Velocities are listed in units of $10^{-3}$.  Values of
  $\sigma$ in a given model indicate the width of Gaussian blurring of
  an emission component in units of keV, using the ``gsmooth'' model
  in XSPEC.  All instances of ``K'' indicate the flux normalization of
  a given component.  Please see the text for additional details.}
\vspace{-1.0\baselineskip}
\end{footnotesize}
\end{table}
\medskip

\clearpage

\begin{table}[t]
\caption{Fits to the Third Order Spectrum of 4U 1630$-$472}
\begin{footnotesize}
\begin{center}
\begin{tabular}{lllllll}
\tableline
Model & Notes & Zone 1 & Zone 2 & Zone 3 & Continuum & $\chi^{2}/\nu$ \\
\tableline
Example & 3 zones, emis. incl., Gauss. blurring & ${\rm N}_{1}$ & ${\rm N}_{2}$ & ${\rm N}_{3}$ & kT (keV) & $500/500=1$ \\
 ~      &         ~         &  ${\rm log}(\xi_{1})$    & ${\rm log}(\xi_{2})$     & ${\rm log}(\xi_{3})$     & ${\rm K}_{disk}$ &   ~       \\
 ~      &         ~         &  $v_{abs, 1}/c$  & $v_{abs, 2}/c$   & $v_{abs, 3}/c$  &  $\Gamma$       &    ~       \\
 ~      &         ~         &  $v_{emis, 1}/c$ & $v_{emis, 2}/c$  & $v_{emis, 3}/c$ &   ${\rm K}_{pow}$ &     ~      \\
 ~      &         ~         &  $\sigma_{emis, 1}$ & $\sigma_{emis, 2}$ & $\sigma_{emis, 2}$      & ~  &    ~  \\
 ~      &         ~         &  ${\rm K}_{emis, 1}$ &  ${\rm K}_{emis, 2}$ &  ${\rm K}_{emis, 3}$ & ~  &  ~    \\

\tableline
1630-3a      & 2 zones, emis. incl., Gauss. blurring & $50^{+10}_{-20}$ & $2^{+1}_{-1}$ & -- & 1.45(5) & $1061/1101 = 0.963$ \\
 ~      &         ~         &  $4.3(2)$    & $4.2(2)$     & --     & $110(20)$ &   ~       \\
 ~      &         ~         &  $-1.0(3)$  & $-7.0_{-2.0}^{+3.0}$   & --  &  --       &    ~       \\
 ~      &         ~         &  $0.2^{+0.2}_{-0.2}$ & $0.7^{0.1}_{-0.7}$  & -- &  -- &     ~      \\
 ~      &         ~         &  $0.16(4)$ & $0.02^{+0.01}_{-0.02}$ & --      & ~  &    ~  \\
 ~      &         ~         &  $0.10^{+0.04}$ &  $0.1(1)$ &  -- & ~  &  ~    \\
\tableline
1630-3b      & 2 zones, emis. incl., no blurring & $60$ & $3$ & -- & 1.45 & $1077/1103 = 0.963$ \\
 ~      &         ~         &  $4.7$    & $3.5$     & --     & 105 &   ~       \\
 ~      &         ~         &  $-0.7$  & $-8.0$   & --  &  --       &    ~       \\
 ~      &         ~         & 0.0  & 3.5  & -- &  -- &     ~      \\
 ~      &         ~         &  -- & -- & --      & ~  &    ~  \\
 ~      &         ~         &  $0.1$ &  $0.1$ &  -- & ~  &  ~    \\
\tableline

1630-3c      & 1 zones, emis. incl., no blurring & $60$ & -- & -- & 1.48 & $1088/1108 = 0.963$ \\
 ~      &         ~         &  $4.6$    & --     & --     & 92 &   ~       \\
 ~      &         ~         &  $-0.9$  & --  & --  &  --       &    ~       \\
 ~      &         ~         & 0.0  & 3.5  & -- &  -- &     ~      \\
 ~      &         ~         &  -- & -- & --      & ~  &    ~  \\
 ~      &         ~         &  $0.1$ &  -- &  -- & ~  &  ~    \\
\tableline
\end{tabular}
\vspace*{\baselineskip}~\\ \end{center} 
\tablecomments{The table above details fits made to the third-order
  spectrum of 4U~1630$-$472 in the 5.0--8.0~keV band, using a
  high-resolution XSTAR photoionization grid.  The ``example'' model explains the parameter values
  listed for subsequent models.  Where ``--'' appears, the component
  was not included in the model.  Asterisks indicate that the
  parameter was frozen to the indicated value in the model.  Errors
  are quoted only for the best model(s).  The hydrogen equivalent
  column density, ${\rm N}$, is given in units of $10^{22}~ {\rm
    cm}^{-2}$ for each absorption/emission zone.  The ionization
  parameter ($\xi = L/nr^{2}$) for the emission/absorption zone is
  quoted in log format for clarity.  When emission components are
  included in a model, the values of ${\rm N}$ and $\xi$ for each zone
  were fixed to same parameters in the absorption component.  Negative
  velocity shifts indicate blue-shifts; positive velocity indicate
  red-shifts.  Velocities are listed in units of $10^{-3}$.  Values of
  $\sigma$ in a given model indicate the width of Gaussian blurring of
  an emission component in units of keV, using the ``gsmooth'' model
  in XSPEC.  All instances of ``K'' indicate the flux normalization of
  a given component.  Please see the text for additional details.}
\vspace{-1.0\baselineskip}
\end{footnotesize}
\end{table}
\medskip

\clearpage

\begin{table}[t]
\caption{Fits to the Third Order Spectrum of H~1743$-$322}
\begin{footnotesize}
\begin{center}
\begin{tabular}{lllllll}
\tableline
Model & Notes & Zone 1 & Zone 2 & Zone 3 & Continuum & $\chi^{2}/\nu$ \\
\tableline
Example & 3 zones, emis. incl., Gauss. blurring & ${\rm N}_{1}$ & ${\rm N}_{2}$ & ${\rm N}_{3}$ & kT (keV) & $500/500=1$ \\
 ~      &         ~         &  ${\rm log}(\xi_{1})$    & ${\rm log}(\xi_{2})$     & ${\rm log}(\xi_{3})$     & ${\rm K}_{disk}$ &   ~       \\
 ~      &         ~         &  $v_{abs, 1}/c$  & $v_{abs, 2}/c$   & $v_{abs, 3}/c$  &  $\Gamma$       &    ~       \\
 ~      &         ~         &  $v_{emis, 1}/c$ & $v_{emis, 2}/c$  & $v_{emis, 3}/c$ &   ${\rm K}_{pow}$ &     ~      \\
 ~      &         ~         &  $\sigma_{emis, 1}$ & $\sigma_{emis, 2}$ & $\sigma_{emis, 2}$      & ~  &    ~  \\
 ~      &         ~         &  ${\rm K}_{emis, 1}$ &  ${\rm K}_{emis, 2}$ &  ${\rm K}_{emis, 3}$ & ~  &  ~    \\

\tableline
1743-3a      & 2 zones, emis. incl., Gauss. blurring & $8_{-5}^{ }$ & $24^{+6}_{-16}$ & -- & 0.8(1) & $1112/1100 = 1.011$ \\
 ~      &         ~         &  $4.7(2)$    & $6_{-1}$     & --     & $6000(3000)$ &   ~       \\
 ~      &         ~         &  $0.0^{+0.1}$  & $-4(1)$   & 2.4*  &  --       &    ~       \\
 ~      &         ~         &  $0.0^{+0.1}$ & $0.0^{+0.1}$  & 5.6(5) &  -- &     ~      \\
 ~      &         ~         &  $0.0^{+0.1}$ & $0.02^{+0.01}_{-0.02}$ & --      & ~  &    ~  \\
 ~      &         ~         &  $2.2_{-1.4}^{+1.0}$ &  $2.2_{-1.6}^{+7.8}$ &  -- & ~  &  ~    \\
\tableline
1743-3b      & 2 zones, emis. incl., no blurring & $8.3$ & $25$ & -- & 0.81 & $1112/1102 = 1.009$ \\
 ~      &         ~         &  $4.7$    & $6.0$     & --     & 5900 &   ~       \\
 ~      &         ~         &  $0.0$  & $-4.3$   & 2.4*  &  --       &    ~       \\
 ~      &         ~         & 0.0  & 0.01  & -- & 5.7 &     ~      \\
 ~      &         ~         &  -- & -- & --      & ~  &    ~  \\
 ~      &         ~         &  $2.3$ &  $0.1$ &  -- & ~  &  ~    \\
\tableline
1743-3c      & 2 zones, no emis., no blurring & $1.4$ & $25.0$ & -- & 0.81 & $1117/1106 = 1.010$ \\
 ~      &         ~         &  $4.5$ & $6.0$    & --     & 5500 &   ~       \\
 ~      &         ~         &  $0.0$  & $-4.3$  & --  &  2.4*       &    ~       \\
 ~      &         ~         & --  & --  & -- & 5.6 &     ~      \\
 ~      &         ~         &  -- & -- & --      & ~  &    ~  \\
 ~      &         ~         &  -- &  -- &  -- & ~  &  ~    \\
\tableline
1743-3d      & 1 zones, no emis., no blurring & $15.3$ & -- & -- & 0.83 & $1124/1109 = 1.014$ \\
 ~      &         ~         &  $5.4$ & --    & --     & 5100 &   ~       \\
 ~      &         ~         &  $-3.9$  & --  & --  &  2.4*       &    ~       \\
 ~      &         ~         & --  & --  & -- &  5.6 &     ~      \\
 ~      &         ~         &  -- & -- & --      & ~  &    ~  \\
 ~      &         ~         &  -- &  -- &  -- & ~  &  ~    \\
\tableline
\end{tabular}
\vspace*{\baselineskip}~\\ \end{center} 
\tablecomments{The table above details fits made to the third-order
  spectrum of H~1743$-$322 in the 5.0--8.0~keV band, using a
  high-resolution XSTAR photoionization grid.  The ``example'' model explains the parameter values
  listed for subsequent models.  Where ``--'' appears, the component
  was not included in the model.  Asterisks indicate that the
  parameter was frozen to the indicated value in the model.  Errors
  are quoted only for the best model(s).  The hydrogen equivalent
  column density, ${\rm N}$, is given in units of $10^{22}~ {\rm
    cm}^{-2}$ for each absorption/emission zone.  The ionization
  parameter ($\xi = L/nr^{2}$) for the emission/absorption zone is
  quoted in log format for clarity.  When emission components are
  included in a model, the values of ${\rm N}$ and $\xi$ for each zone
  were fixed to same parameters in the absorption component.  Negative
  velocity shifts indicate blue-shifts; positive velocity indicate
  red-shifts.  Velocities are listed in units of $10^{-3}$.  Values of
  $\sigma$ in a given model indicate the width of Gaussian blurring of
  an emission component in units of keV, using the ``gsmooth'' model
  in XSPEC.  All instances of ``K'' indicate the flux normalization of
  a given component.  Please see the text for additional details.}
\vspace{-1.0\baselineskip}
\end{footnotesize}
\end{table}
\medskip

\clearpage

\begin{table}[t]
\caption{Fits to the First Order Spectrum of GRO J1655$-$40}
\begin{footnotesize}
\begin{center}
\begin{tabular}{lllllll}
\tableline
Model & Notes & Zone 1 & Zone 2 & Zone 3 & Continuum & $\chi^{2}/\nu$ \\
\tableline
Example & 3 zones, emis. incl., Gauss. blurring & ${\rm N}_{1}$ & ${\rm N}_{2}$ & ${\rm N}_{3}$ & kT (keV) & $500/500=1$ \\
 ~      &         ~         &  ${\rm log}(\xi_{1})$    & ${\rm log}(\xi_{2})$     & ${\rm log}(\xi_{3})$     & ${\rm K}_{disk}$ &   ~       \\
 ~      &         ~         &  $v_{abs, 1}/c$  & $v_{abs, 2}/c$   & $v_{abs, 3}/c$  &  $\Gamma$       &    ~       \\
 ~      &         ~         &  $v_{emis, 1}/c$ & $v_{emis, 2}/c$  & $v_{emis, 3}/c$ &   ${\rm K}_{pow}$ &     ~      \\
 ~      &         ~         &  $\sigma_{emis, 1}$ & $\sigma_{emis, 2}$ & $\sigma_{emis, 2}$      & ~  &    ~  \\
 ~      &         ~         &  ${\rm K}_{emis, 1}$ &  ${\rm K}_{emis, 2}$ &  ${\rm K}_{emis, 3}$ & ~  &  ~    \\

\tableline
1655-1a      & 3 zones, emis. incl., Gauss. blurring & $59^{+1}_{-3}$ & $8.2^{+0.5}_{-1.2}$ & $4.2^{+0.4}_{-0.7}$ & 1.22(1) & $634/477 = 1.329$ \\
 ~      &         ~         &  $4.72(4)$    & $4.53(3)$     & $4.95(5)$     & 930(30) &   ~       \\
 ~      &         ~         &  $-1.5(1)$  & $-6.4(4)$   & $-11.8*$  &  3.5*       &    ~       \\
 ~      &         ~         &  $0^{+0.1}$ & $0.0^{+0.1}$  & $0.0^{+0.1}$ &  $4_{-3}^{+10}$ &     ~      \\
 ~      &         ~         &  $0.25(5)$ & $0.22(4)$ & $0.018(4)$      & $0.12^{+0.11}_{-0.04}$ &    ~  \\
 ~      &         ~         &  $0.10^{0.01}$ &  $2.5(5)$ &  $10_{-2}$ & $0.5^{+0.6}_{-0.4}$  &  ~    \\
\tableline
1655-1b     & 3 zones, emis. incl., no blurring & $58$ & $8.2$ & $3.2$ & 1.25 & $928/480 = 1.934$ \\
 ~      &         ~         &  $4.7$    & $5.1$     & $4.9$     & 800 &   ~       \\
 ~      &         ~         &  $-1.3$  & $-6.7$   & $-11.8*$  &  3.5*       &    ~       \\
 ~      &         ~         &  $0$ & $0.0$  & $0$ &   0 &     ~      \\
 ~      &         ~         &  -- & -- & --      & 0.12  &    ~  \\
 ~      &         ~         &  $0.1$ &  $5.0$ &  $0.1$ & 0.5  &  ~    \\
\tableline
1655-1c     & 3 zones, no emis., no blurring & $19$ & $0.4$ & $0.3$ & 1.27 & $963.0/484 = 1.990$ \\
 ~      &         ~         &  $4.3$    & $4.7$     & $3.3$     & 741 &   ~       \\
 ~      &         ~         &  $-2.3$  & $-8.4$  & $-11.8*$  &  3.5*       &    ~       \\
 ~      &         ~         &  -- & --  & -- &   0 &     ~      \\
 ~      &         ~         &  -- & -- & --      & 0.12  &    ~  \\
 ~      &         ~         &  -- &  -- &  -- & 1.8  &  ~    \\
\tableline
1655-1d      & 2 zones, emis. incl., Gauss. blurring & $54$ & -- & $4.4$ & 1.22(1) & $732/481 = 1.521$ \\
 ~      &         ~         &  $4.6$    & --     & $4.7$     & 880 &   ~       \\
 ~      &         ~         &  $-2.2$  & --   & $-11.8*$  &  3.5*       &    ~       \\
 ~      &         ~         &  $0$ & --  & $0$ &   0 &     ~      \\
 ~      &         ~         &  $0.24$ & -- & $0.02$      & 0.09  &    ~  \\
 ~      &         ~         &  $0.25$ & -- &  $3.7$ & 0.05  &  ~    \\
\tableline
1655-1e      & 2 zones, emis. incl., no blurring & $56$ & -- & $0.8$ & 1.25(1) & $732/481 = 1.521$ \\
 ~      &         ~         &  $4.7$    & --     & $5.3$     & 810 &   ~       \\
 ~      &         ~         &  $-1.7$  & --   & $-11.8*$  &  3.5*       &    ~       \\
 ~      &         ~         &  $0$ & --  & $0$ &   0 &     ~      \\
 ~      &         ~         &  -- & -- &  --     & 0.12  &    ~  \\
 ~      &         ~         &  $0.1$ & -- &  $6.2$ & 2.0  &  ~    \\
\tableline
1655-1f     & 2 zones, no emis., no blurring & $35$ & -- & $3.8$ & 1.27 & $995/486 = 2.043$ \\
 ~      &         ~         &  $4.5$    & --     & $4.8$     & 740 &   ~       \\
 ~      &         ~         &  $-2.7$  & --   & $-11.8*$  &  3.5*       &    ~       \\
 ~      &         ~         &  -- & --  & -- &   0.001 &     ~      \\
 ~      &         ~         &  -- & -- & --     & 0.3  &    ~  \\
 ~      &         ~         &  -- & -- & -- & 3.7  &  ~    \\
\tableline
\end{tabular}
\vspace*{\baselineskip}~\\ \end{center} 
\tablecomments{The table above details fits made to the first-order
  spectrum of GRO J1655$-$40 in the 5.0--10.0~keV band, using a
  high-resolution XSTAR photoionization grid.  The ``example'' model explains the parameter values
  listed for subsequent models.  Where ``--'' appears, the component
  was not included in the model.  Asterisks indicate that the
  parameter was frozen to the indicated value in the model.  Errors
  are quoted only for the best model(s).  The hydrogen equivalent
  column density, ${\rm N}$, is given in units of $10^{22}~ {\rm
    cm}^{-2}$ for each absorption/emission zone.  The ionization
  parameter ($\xi = L/nr^{2}$) for the emission/absorption zone is
  quoted in log format for clarity.  When emission components are
  included in a model, the values of ${\rm N}$ and $\xi$ for each zone
  were fixed to same parameters in the absorption component.  Negative
  velocity shifts indicate blue-shifts; positive velocity indicate
  red-shifts.  Velocities are listed in units of $10^{-3}$.  Values of
  $\sigma$ in a given model indicate the width of Gaussian blurring of
  an emission component in units of keV, using the ``gsmooth'' model
  in XSPEC.  All instances of ``K'' indicate the flux normalization of
  a given component.}
\vspace{-1.0\baselineskip}
\end{footnotesize}
\end{table}
\medskip

\begin{table}[t]
\caption{Fits to the First Order Spectrum of GRS 1915$+$105}
\begin{footnotesize}
\begin{center}
\begin{tabular}{lllllll}
\tableline
Model & Notes & Zone 1 & Zone 2 & Zone 3 & Continuum & $\chi^{2}/\nu$ \\
\tableline
Example & 3 zones, emis. incl., Gauss. blurring & ${\rm N}_{1}$ & ${\rm N}_{2}$ & ${\rm N}_{3}$ & kT (keV) & $500/500=1$ \\
 ~      &         ~         &  ${\rm log}(\xi_{1})$    & ${\rm log}(\xi_{2})$     & ${\rm log}(\xi_{3})$     & ${\rm K}_{disk}$ &   ~       \\
 ~      &         ~         &  $v_{abs, 1}/c$  & $v_{abs, 2}/c$   & $v_{abs, 3}/c$  &  $\Gamma$       &    ~       \\
 ~      &         ~         &  $v_{emis, 1}/c$ & $v_{emis, 2}/c$  & $v_{emis, 3}/c$ &   ${\rm K}_{pow}$ &     ~      \\
 ~      &         ~         &  $\sigma_{emis, 1}$ & $\sigma_{emis, 2}$ & $\sigma_{emis, 2}$      & ~  &    ~  \\
 ~      &         ~         &  ${\rm K}_{emis, 1}$ &  ${\rm K}_{emis, 2}$ &  ${\rm K}_{emis, 3}$ & ~  &  ~    \\

\tableline
1915-1a      & 3 zones, emis. incl., Gauss. blurring & $42(4)$ & $60_{-10}$ & $0.7_{-0.2}^{+0.1}$ & 1.90(1) & $757/476 = 1.591$ \\
 ~      &         ~         &  $3.76(3)$    & $5.04^{+0.06}_{-0.2}$     & $3.82(4)$     & 68(2) &   ~       \\
 ~      &         ~         &  $-0.5(1)$  & $-1.7(3)$   & $-7.2(6)$  &  3.0*       &    ~       \\ 
~      &         ~         &  $0.0^{+0.1}$ & $0.0^{+0.1}$  & $0.0^{+0.1}$ &  $0.0^{+0.2}$ &  ~     \\
 ~      &         ~         &  $0.29_{-0.01}$ & $0.02(1)$ & $0.07(1)$      & $0.11^{+0.05}_{-0.03}$  &    ~  \\
 ~      &         ~         &  $0.10^{+0.02}$ &  $0.28^{+0.06}_{-0.09}$ &  $1.7^{+0.4}_{-0.2}$ & $1.2(4)$  &  ~    \\
\tableline
1915-1b      & 3 zones, emis. incl., no blurring & $47$ & $60$ & $0.6$ & 1.92 & $1003/479 = 2.095$ \\
 ~      &         ~         &  $3.9$    & $4.5$     & $6.0$     & 66 &   ~       \\
 ~      &         ~         &  $0.0$  & $-0.9$   & $-65$  &  3.0*       &    ~       \\
 ~      &         ~         &  $0.0$ & $0.0$  & $0.0$ &  0.0 &     ~      \\
 ~      &         ~         &  -- & -- & --      & 0.3  &    ~  \\
 ~      &         ~         &  $0.10$ &  $1.2$ &  $20$ & 6.6  &  ~    \\
\tableline
1915-1c      & 3 zones, no emis., no blurring & $18$ & $0.3$ & $0.3$ & 1.94 & $1040/482 = 2.153$ \\
 ~      &         ~         &  $4.0$    & $3.2$     & $6.0$     & 63 &   ~       \\
 ~      &         ~         &  $-1.0$  & $-19$   & $-99$  &  3.0*       &    ~       \\
 ~      &         ~         &  -- & --  & -- &  0.0 &     ~      \\
 ~      &         ~         &  -- & -- & --      & 0.3  &    ~  \\
 ~      &         ~         &  -- & -- & -- & 7.2  &  ~    \\
\tableline
1915-1d      & 2 zones, emis. incl., Gauss. blurring & $44$ & 60 & -- & 1.91 & $829/481 = 1.724$ \\
 ~      &         ~         &  $4.0$    & 5.3     & --     & 67 &   ~       \\
 ~      &         ~         &  $-0.6$  & $-2.3$   & --  &  3.0*       &    ~       \\
 ~      &         ~         &  $0.0$ & $0.0$  & -- &   0 &     ~      \\
 ~      &         ~         &  $0.10$ & $0.02$ & --      & 0.27  &    ~  \\
 ~      &         ~         &  $0.10$ & $0.8$  & -- & 5.2  &  ~    \\
\tableline
1915-1e      & 2 zones, emis. incl., no blurring & $43$ & 49 & -- & 1.92 & $1007/483 = 2.084$ \\
 ~      &         ~         &  $3.9$    & 4.4     & --     & 66 &   ~       \\
 ~      &         ~         &  $0.0$  & $-1.0$   & --  &  3.0*       &    ~       \\
 ~      &         ~         &  $0.0$ & $0.0$  & -- &   0 &     ~      \\
 ~      &         ~         &  -- & -- & --      & 0.3 &    ~  \\
 ~      &         ~         &  $0.1$ & $0.1$  & -- & 7.0  &  ~    \\
\tableline
1915-1f      & 2 zones, no emis., no blurring & $4.0$ & 41 & -- & 1.92 & $1203/486 = 2.475$ \\
 ~      &         ~         &  $3.9$    & 4.8     & --     & 66 &   ~       \\
 ~      &         ~         &  $0.0$  & $-1.6$   & --  &  3.0*       &    ~       \\
 ~      &         ~         &  -- & --  & -- &   0 &     ~      \\
 ~      &         ~         &  -- & -- & --      & 0.3  &    ~  \\
 ~      &         ~         &  -- & --  & -- & 7.0  &  ~    \\
\tableline
\end{tabular}
\vspace*{\baselineskip}~\\ \end{center} 
\tablecomments{The table above details fits made to the first-order
  spectrum of GRS 1915$+$105 in the 5.0--10.0~keV band, using a
  high-resolution XSTAR photoionization grid.  The ``example'' model explains the parameter values
  listed for subsequent models.  Where ``--'' appears, the component
  was not included in the model.  Asterisks indicate that the
  parameter was frozen to the indicated value in the model.  Errors
  are quoted only for the best model(s).  The hydrogen equivalent
  column density, ${\rm N}$, is given in units of $10^{22}~ {\rm
    cm}^{-2}$ for each absorption/emission zone.  The ionization
  parameter ($\xi = L/nr^{2}$) for the emission/absorption zone is
  quoted in log format for clarity.  When emission components are
  included in a model, the values of ${\rm N}$ and $\xi$ for each zone
  were fixed to same parameters in the absorption component.  Negative
  velocity shifts indicate blue-shifts; positive velocity indicate
  red-shifts.  Velocities are listed in units of $10^{-3}$.  Values of
  $\sigma$ in a given model indicate the width of Gaussian blurring of
  an emission component in units of keV, using the ``gsmooth'' model
  in XSPEC.  All instances of ``K'' indicate the flux normalization of
  a given component.  Please see the text for additional details.}
\vspace{-1.0\baselineskip}
\end{footnotesize}
\end{table}
\medskip

\clearpage

\begin{table}[t]
\caption{Fits to the First Order Spectrum of 4U 1630$-$472}
\begin{footnotesize}
\begin{center}
\begin{tabular}{lllllll}
\tableline
Model & Notes & Zone 1 & Zone 2 & Zone 3 & Continuum & $\chi^{2}/\nu$ \\
\tableline
Example & 3 zones, emis. incl., Gauss. blurring & ${\rm N}_{1}$ & ${\rm N}_{2}$ & ${\rm N}_{3}$ & kT (keV) & $500/500=1$ \\
 ~      &         ~         &  ${\rm log}(\xi_{1})$    & ${\rm log}(\xi_{2})$     & ${\rm log}(\xi_{3})$     & ${\rm K}_{disk}$ &   ~       \\
 ~      &         ~         &  $v_{abs, 1}/c$  & $v_{abs, 2}/c$   & $v_{abs, 3}/c$  &  $\Gamma$       &    ~       \\
 ~      &         ~         &  $v_{emis, 1}/c$ & $v_{emis, 2}/c$  & $v_{emis, 3}/c$ &   ${\rm K}_{pow}$ &     ~      \\
 ~      &         ~         &  $\sigma_{emis, 1}$ & $\sigma_{emis, 2}$ & $\sigma_{emis, 2}$      & ~  &    ~  \\
 ~      &         ~         &  ${\rm K}_{emis, 1}$ &  ${\rm K}_{emis, 2}$ &  ${\rm K}_{emis, 3}$ & ~  &  ~    \\

\tableline
1630-1a      & 2 zones, emis. incl., Gauss. blurring & $22(2)$ & $7^{+3}_{-2}$ & -- & 1.49(1) & $518/481 = 1.074$ \\
 ~      &         ~         &  $4.14(2)$    & $4.6^{+0.3}_{-0.1}$     & --     & $126(5)$ &   ~       \\
 ~      &         ~         &  $-0.7_{-0.3}^{+0.4}$  & $-6.0(2)$   & --  &  --       &    ~       \\
 ~      &         ~         &  $0.0^{+0.1}$ & $0.0^{+0.1}$  & -- &  -- &     ~      \\
 ~      &         ~         &  $0.29_{-0.12}^{+0.03}$ & $0.0^{+0.05}$ & --      & $0.20_{-0.08}$  &    ~  \\
 ~      &         ~         &  $0.11^{+0.04}$ &  $0.3(2)$ &  -- & $0.9(4)$ &  ~    \\
\tableline
1630-1b      & 2 zones, emis. incl., no blurring & $8.2$ & $31.0$ & -- & 1.49 & $573/483 = 1.185$ \\
 ~      &         ~         &  $4.0$    & $4.6$     & --     & $127$ &   ~       \\
 ~      &         ~         &  $0$  & $0$   & --  &  --       &    ~       \\
 ~      &         ~         &  $0.0$ & $0.0$  & -- &  -- &     ~      \\
 ~      &         ~         &  -- & --  & --      & 0.2  &    ~  \\
 ~      &         ~         &  $0.1$ &  $0.1$ &  -- & 1.2  &  ~    \\
\tableline
1630-1c      & 2 zones, no emis., no blurring & $6.6$ & $20.2$ & -- & 1.51 & $528/485 = 1.088$ \\
 ~      &         ~         &  $4.1$    & $4.5$     & --     & $119$ &   ~       \\
 ~      &         ~         &  $0.0$  & $0.0$   & --  &  --       &    ~       \\
 ~      &         ~         &  -- &  --  & -- &  -- &     ~      \\
 ~      &         ~         &  -- & -- & --      & 0.2  &    ~  \\
 ~      &         ~         &  -- & -- &  -- & 1.4  &  ~    \\
\tableline
1630-1d      & 1 zone, no emis., no blurring & $22.2$ & -- & -- & 1.49 & $529/488 = 1.084$ \\
 ~      &         ~         &  $4.17$    & --    & --     & $126$ &   ~       \\
 ~      &         ~         &  $-1.2$  & --   & --  &  --       &    ~       \\
 ~      &         ~         &  -- & --  & -- &  -- &     ~      \\
 ~      &         ~         &  -- & -- & --      & ~  &    ~  \\
 ~      &         ~         &  -- &  -- &  -- & ~  &  ~    \\

\tableline
\end{tabular}
\vspace*{\baselineskip}~\\ \end{center} 
\tablecomments{The table above details fits made to the first-order
  spectrum of 4U 1630$-$472 in the 5.0--10.0~keV band, using a
  high-resolution XSTAR photoionization grid.  The ``example'' model explains the parameter values
  listed for subsequent models.  Where ``--'' appears, the component
  was not included in the model.  Asterisks indicate that the
  parameter was frozen to the indicated value in the model.  Errors
  are quoted only for the best model(s).  The hydrogen equivalent
  column density, ${\rm N}$, is given in units of $10^{22}~ {\rm
    cm}^{-2}$ for each absorption/emission zone.  The ionization
  parameter ($\xi = L/nr^{2}$) for the emission/absorption zone is
  quoted in log format for clarity.  When emission components are
  included in a model, the values of ${\rm N}$ and $\xi$ for each zone
  were fixed to same parameters in the absorption component.  Negative
  velocity shifts indicate blue-shifts; positive velocity indicate
  red-shifts.  Velocities are listed in units of $10^{-3}$.  Values of
  $\sigma$ in a given model indicate the width of Gaussian blurring of
  an emission component in units of keV, using the ``gsmooth'' model
  in XSPEC.  All instances of ``K'' indicate the flux normalization of
  a given component.  Please see the text for additional details.}
\vspace{-1.0\baselineskip}
\end{footnotesize}
\end{table}
\medskip

\clearpage

\begin{table}[t]
\caption{Fits to the First Order Spectrum of H 1743$-$322}
\begin{footnotesize}
\begin{center}
\begin{tabular}{lllllll}
\tableline
Model & Notes & Zone 1 & Zone 2 & Zone 3 & Continuum & $\chi^{2}/\nu$ \\
\tableline
Example & 3 zones, emis. incl., Gauss. blurring & ${\rm N}_{1}$ & ${\rm N}_{2}$ & ${\rm N}_{3}$ & kT (keV) & $500/500=1$ \\
 ~      &         ~         &  ${\rm log}(\xi_{1})$    & ${\rm log}(\xi_{2})$     & ${\rm log}(\xi_{3})$     & ${\rm K}_{disk}$ &   ~       \\
 ~      &         ~         &  $v_{abs, 1}/c$  & $v_{abs, 2}/c$   & $v_{abs, 3}/c$  &  $\Gamma$       &    ~       \\
 ~      &         ~         &  $v_{emis, 1}/c$ & $v_{emis, 2}/c$  & $v_{emis, 3}/c$ &   ${\rm K}_{pow}$ &     ~      \\
 ~      &         ~         &  $\sigma_{emis, 1}$ & $\sigma_{emis, 2}$ & $\sigma_{emis, 2}$      & ~  &    ~  \\
 ~      &         ~         &  ${\rm K}_{emis, 1}$ &  ${\rm K}_{emis, 2}$ &  ${\rm K}_{emis, 3}$ & ~  &  ~    \\

\tableline
1743-1a      & 2 zones, emis. incl., Gauss. blurring & $6.1(6)$ & $14_{-8}^{+3}$ & -- & 1.10(1) & $723/478 = 1.513$ \\
 ~      &         ~         &  $4.57(3)$    & $6.0_{-0.5}$     & --     & $1090(50)$ &   ~       \\
 ~      &         ~         &  $-1.0(3)$  & $-8.6(9)$   & --  &  2.4*       &    ~       \\
 ~      &         ~         &  $50(20)$ & $5^{+15}$  & -- &  2.45(8) &     ~      \\
 ~      &         ~         &  $0.08_{-0.08}^{+0.08}$ & $0.20_{-0.01}$ & --      & $0.2_{-0.02}$  &    ~  \\
 ~      &         ~         &  $0.5_{-0.4}^{+0.7}$ &  $10_{-2}$ &  -- & 2.3(3)  &  ~    \\
\tableline
1743-1b      & 2 zones, emis. incl., no blurring & $5.7$ & $13.2$ & -- & 1.10 & $728/480 = 1.518$ \\
 ~      &         ~         &  $4.58$    & $6.0$     & --     & $1195$ &   ~       \\
 ~      &         ~         &  $-1.0(3)$  & $-8.5$   & --  &  2.4*       &    ~       \\
 ~      &         ~         &  $0.0$ & $0.0$  & -- &  2.8 &     ~      \\
 ~      &         ~         &  -- & -- & --      & 0.2  &    ~  \\
 ~      &         ~         &  $0.12$ &  $0.68$ &  -- & 2.1  &  ~    \\
\tableline
1743-1c      & 2 zones, no emis., no blurring & $5.7$ & $13.2$ & -- & 1.10 & $739/484 = 1.527$ \\
 ~      &         ~         &  $4.58$    & $6.0$     & --     & $1215$ &   ~       \\
 ~      &         ~         &  $-1.0$  & $-8.5$   & --  &  2.4*       &    ~       \\
 ~      &         ~         &  -- & --  & -- &  2.8 &     ~      \\
 ~      &         ~         &  -- & -- & --      & 0.2  &    ~  \\
 ~      &         ~         &  -- &  -- &  -- & 2.1 &  ~    \\
\tableline
1743-1d     & 1 zones, no emis., no blurring & $6.0$ & -- & -- & 1.09 & $742/487 = 1.579$ \\
 ~      &         ~         &  $4.59$    & --     & --     & $1225$ &   ~       \\
 ~      &         ~         &  $-1.3$  & --   & --  &  2.4*       &    ~       \\
 ~      &         ~         &  -- & --  & -- &  2.8 &     ~      \\
 ~      &         ~         &  -- & -- & --      & 0.2  &    ~  \\
 ~      &         ~         &  -- &  -- &  -- & 2.3  &  ~    \\
\tableline
\end{tabular}
\vspace*{\baselineskip}~\\ \end{center} 
\tablecomments{The table above details fits made to the first-order
  spectrum of H 1743$-$322 in the 5.0--10.0~keV band, using a
  high-resolution XSTAR photoionization grid.  The ``example'' model explains the parameter values
  listed for subsequent models.  Where ``--'' appears, the component
  was not included in the model.  Asterisks indicate that the
  parameter was frozen to the indicated value in the model.  Errors
  are quoted only for the best model(s).  The hydrogen equivalent
  column density, ${\rm N}$, is given in units of $10^{22}~ {\rm
    cm}^{-2}$ for each absorption/emission zone.  The ionization
  parameter ($\xi = L/nr^{2}$) for the emission/absorption zone is
  quoted in log format for clarity.  When emission components are
  included in a model, the values of ${\rm N}$ and $\xi$ for each zone
  were fixed to same parameters in the absorption component.  Negative
  velocity shifts indicate blue-shifts; positive velocity indicate
  red-shifts.  Velocities are listed in units of $10^{-3}$.  Values of
  $\sigma$ in a given model indicate the width of Gaussian blurring of
  an emission component in units of keV, using the ``gsmooth'' model
  in XSPEC.  All instances of ``K'' indicate the flux normalization of
  a given component.  Please see the text for additional details.}
\vspace{-1.0\baselineskip}
\end{footnotesize}
\end{table}
\medskip

\clearpage

\begin{table}[t]
\caption{Radius-focused Fits to First Order Spectra}
\begin{footnotesize}
\begin{center}
\begin{tabular}{lllllll}
\tableline
Model & Notes & Zone 1 & Zone 2 & Zone 3 & Continuum & $\chi^{2}/\nu$ \\
\tableline
Example & 3 zones, emis. incl., Gauss. blurring & ${\rm N}_{1}$ & ${\rm N}_{2}$ & ${\rm N}_{3}$ & kT (keV) & $500/500=1$ \\
 ~      &         ~         &  ${\rm log}(\xi_{1})$    & ${\rm log}(\xi_{2})$     & ${\rm log}(\xi_{3})$     & ${\rm K}_{disk}$ &   ~       \\
 ~      &         ~         &  $v_{abs, 1}/c$  & $v_{abs, 2}/c$   & $v_{abs, 3}/c$  &  $\Gamma$       &    ~       \\
 ~      &         ~         &  $R_{in,1}$ & $R_{in,2}$  & $R_{in,3}$ &   ${\rm K}_{pow}$ &     ~      \\
 ~      &         ~         &  $\theta_{1}$ & $\theta_{2}$ & $\theta_{3}$      & $\sigma$ (keV)  &    ~  \\
 ~      &         ~         &  ${\rm K}_{emis, 1}$ &  ${\rm K}_{emis, 2}$ &  ${\rm K}_{emis, 3}$ & $K_{gauss}$  &  ~    \\
\tableline
1655-r1 & 3 zones, emis. incl., linked blurring & $49(2)$ & $7.6(8)$ & $8^{+2}_{-1}$ & $1.22(1)$ & $710/477 = 1.490$ \\
 ~      &         ~         &  $4.59(3)$    & $4.65(4)$     & $5.0^{+0.3}_{-0.1}$     & $940(10)$ &   ~       \\
 ~      &         ~         &  $-2.2(1)$  & $-7.8(5)$   & $-11.8*$  &  $3.5*$       &    ~       \\
 ~      &         ~         &  $1600^{+300}_{-100}$ &   $1600_{-100}^{+300}$  & $1600^{+300}_{-100}$ &   $0.7^{+1.5}_{-0.7}$ &     ~      \\
 ~      &         ~         &  $70-85$ & $70-85$ & $70-85$     & $0.12(2)$  &    ~  \\
 ~      &         ~         &  $0.10^{+0.05}$ &  $2.0(7)$ &  $10_{-1}$ & $2.4(4)$  &  ~    \\
\tableline
1655-r2 & 3 zones, emis. incl., indep. blurring & $55(3)$ & $8.3_{-0.5}^{+1.7}$ & $4.7(7)$ & $1.22(1)$ & $671/475 = 1.414$ \\
 ~      &         ~         &  $4.67(5)$    & $4.62(2)$     & $4.95(5)$     & $920(20)$ &   ~       \\
 ~      &         ~         &  $-2.0(2)$  & $-7.0(5)$   & $-11.8*$  &  $3.5*$       &    ~       \\
 ~      &         ~         &  $500^{+400}$ & $1100^{+200}_{-200}$  & $10,000_{-1000}$ &   $0.0^{+0.5}$ &     ~      \\
 ~      &         ~         &  $70^{+3}$ & $70^{+3}$ & $70^{+3}$      & $0.12(2)$  &    ~  \\
 ~      &         ~         &  $0.10^{+0.01}$ &  $2.5(5)$ &  $10_{-1}$ & $2.0(4)$  &  ~    \\
\tableline
1915-r1 & 3 zones, emis. incl., linked blurring & $40(7)$ & $4(1)$ & $60_{-30}$ & $1.90(1)$ & $793/477 = 1.662$ \\
 ~      &         ~         &  $4.10(5)$    & $3.78(5)$     & $5.9^{+0.1}_{-0.6}$     & $68(2)$ &   ~       \\
 ~      &         ~         &  $-1.0(1)$  & $-0.5(5)$   & $-9.4(2)$  &  $3.0*$       &    ~       \\
 ~      &         ~         &  $2400_{-400}^{+600}$ &   $2400_{-400}^{+600}$  & $2400_{-400}^{+600}$ &   $0.0^{+0.1}$ &     ~      \\
 ~      &         ~         &  $60-80$ & $60-80$ & $60-80$     & $0.14(3)$  &    ~  \\
 ~      &         ~         &  $0.10^{+0.02}$ &  $2.0(2)$ &  $2.1(3)$ & $3.0(5)$ &  ~    \\
\tableline
1915-r2 & 3 zones, emis. incl., indep. blurring & $40^{+1}_{-5}$ & $4.2(5)$ & $60_{-3}$ & $1.90(1)$ & $787/475 = 1.646$ \\
 ~      &         ~         &  $4.10(5)$    & $3.8(1)$     & $5.9^{+0.1}_{-0.5}$     & $68(2)$ &   ~       \\
 ~      &         ~         &  $-1.0(1)$  & $-0.5(5)$   & $-9.4(2)$  &  $3.0*$       &    ~       \\
 ~      &         ~         &  $1200_{-200}^{+600}$ &   $5500_{-700}^{+1000}$  & $2400_{-400}^{+600}$ &   $0.0^{+0.1}$ &     ~      \\
 ~      &         ~         &  $60-80$ & $60-80$ & $60-80$     & $0.14(3)$  &    ~  \\
 ~      &         ~         &  $0.10^{+0.03}$ &  $2.0(2)$ &  $2.1(3)$ & $2.6(6)$ &  ~    \\

\tableline
1630-r1 & 2 zones, emis. incl., linked blurring & $21^{+2}_{-3}$ & $5.0^{+3}_{-2}$ & -- & $1.48(2)$ & $519/481 = 1.079$ \\
 ~      &         ~         &  $4.12(2)$    & $4.6(2)$    & --     & $130(4)$ &   ~       \\
 ~      &         ~         &  $-0.8(3)$  & $-6(2)$   & --  &  --       &    ~       \\
 ~      &         ~         &  $800^{+500}_{-200}$ &   $800^{+500}_{-200}$  & -- &   -- &     ~      \\
 ~      &         ~         &  $65-80$ & $65-80$ & --     & 0.17(3)  &    ~  \\
 ~      &         ~         &  $0.10^{+0.05}$ &  $0.1^{+2}$ &  -- & 0.9(4)  &  ~    \\
\tableline

\tableline
1743-r1 & 2 zones, emis. incl., linked blurring & $6.1(6)$ & $14^{+2}_{-9}$ & -- & $1.09(1)$ & $725/482 = 1.509$ \\
 ~      &         ~         &  $4.57(3)$    & $6_{-1}$    & --     & $1250(80)$ &   ~       \\
 ~      &         ~         &  $-1.0(3)$  & $-8.7(9)$   & --  &  2.4*      &    ~       \\
 ~      &         ~         &  $1100^{+1100}_{-300}$ &   $1100^{+1100}_{-300}$  & 2.8(1) &   -- &     ~      \\
 ~      &         ~         &  $60-80$ & $60-80$ & --     & $0.20_{-0.03}$  &    ~  \\
 ~      &         ~         &  $0.7(6)$ &  $2_{-1}^{+2}$ &  -- & $2.1(4)$  &  ~    \\
\tableline
\end{tabular}
\vspace*{\baselineskip}~\\ \end{center} 
\tablecomments{In the table above, Gaussian blurring of the (re-)
  emission from the wind has been replaced with more physical
  relativistic blurring using the ``rdblur'' convolution function.
  Within ``rdblur'', an emissivity profile of $r^{-3}$, and an outer
  outer radius of $10^{5}~ {\rm GM/c}^{2}$ was assumed in all cases, for
  simplicity.  Please see the text for more details.}
\vspace{-1.0\baselineskip}
\end{footnotesize}
\end{table}
\medskip

\begin{table}[t]
\caption{Wind Launching and Outflow Parameters}
\begin{footnotesize}
\begin{center}
\begin{tabular}{lllllllll}
Source, zone & $L_{rad}/L_{Edd.}$  & $\dot{M}_{wind}$ & $\dot{M}_{wind, Edd.}$ & $L_{wind}$ & $L_{wind}/L_{edd}$ & $L_{wind}/L_{rad}$ & $r_{phot.}$ & $r_{blur}$ \\
     ~       &       ~         &   ($10^{18}$~g/s)   &      ~             &  ($10^{34}$ erg/s) &   ($10^{-4}$)   &   ($10^{-3}$)    & (GM/c$^{2}$)   &    (GM/c$^{2}$) \\
\tableline
1630-r1, zone 1 & 0.09          &   3(1)                &    0.2(1)       &    0.07(5)          &  0.006(4)   & $0.006(5)$    &  $6100\pm 2200$  &  $800^{+500}_{-200}$                \\ 
1630-r1, zone 2 & 0.09          &   6(3)                &    0.4(2)       &    9(6)      &   0.7(5)         &  $0.8(6)$   &  $3300\pm 1600$  &  $800^{+500}_{-200}$                \\ 
\tableline
1743-r1, zone 1 & 0.15         &   2.1(8)           &    0.14(6)      &   0.09(6)       &  0.007(4)     &  $0.005(4)$      &  $4900\pm 1700$  &  $1100^{+1100}_{-300}$                \\ 
1743-r1, zone 2 & 0.15          &   1.2(6)          &       0.08(4)   &   4(2)      &    0.3(2)        &   $0.2(1)$      &  $1300\pm 900$  &  $1000^{+1100}_{-300}$                \\ 
\tableline
1655-r2, zone 1 & 0.06          &  0.32(8)        &  0.032(8)           &   0.06(2)       &  0.006(2)   &  $0.01(7)$      &  $3100\pm 600$  &  $500^{+400}$                \\ 
1655-r2, zone 2 & 0.06          &  1.3(3)         &   0.13(3)         &    2.8(8)      &     0.32(8)    & $0.6(3)$        &  $3300\pm 600$  &  $1100\pm 200$                \\ 
1655-r2, zone 3 & 0.06          &  1.0(3)          &    0.1(3)       &   6(2)       &    0.006(2)     &  $1.2(7)$    &  $2300\pm 400$  &  $10000_{-1000}$                \\ 
\tableline
1915-r2, zone 1 & 0.61          &  24(6)       &      1.7(5)       &    1.1(3)      &           0.08(2)  &  $0.013(7)$        &  $17,000\pm 3000$  &  $1200^{+600}_{-200}$                \\ 
1915-r2, zone 2 & 0.61          &  26(22)       &      2(2)        &    $0.4_{-0.4}^{0.5}$  &       0.03(3) &  $0.005(5)$     &  $23,000\pm 6000$  &  $5500^{+1000}_{-700}$                \\ 
1915-r2, zone 3 & 0.61          &  5(2)             &     0.3(1)   &    19(7)      &     1.5(6)        &   $0.2(1)$         &  $2400\pm 1300$  &  $2400^{+600}_{-400}$                \\ 
\tableline
\end{tabular}
\vspace*{\baselineskip}~\\ \end{center} 
\tablecomments{Critical wind parameters are listed, using the
  data in Table 2 and Table 11.  Mass outflow rates were calculated
  via $M_{wind} = \Omega \mu m_p v L / \xi$, and values of wind
  kinetic luminosity via $L_{wind} = 0.5 \dot{M} v^{2}$ (where
  $\Omega$ is the filling factor; $\mu$ is the mean atomic weight and
  $\mu = 1.23$ is assumed; $m_{p}$ is the proton mass, $\xi$ is the
  ionization parameter measured via XSTAR grids; and $v$ is the measured blue-shift).  In
  all cases, a volume filling factor of unity is assumed.  Two radii
  are given; $r_{phot}$ is the photoionization radius derived from $r
  = \sqrt(L/n\xi$), and $r_{blur}$ is the radius derived from blurring
  of the photionized re-emission from the wind.  Luminosity
  uncertainties of 50\% are assumed for all sources.
  The values of $r_{phot}$ are used to calculate the mass outflow rate
  and kinetic luminosity.}
\vspace{-1.0\baselineskip}
\end{footnotesize}
\end{table}
\medskip

\end{document}